\documentclass[useAMS,usenatbib,usegraphicx]{mn2e}
\usepackage{graphicx}
\usepackage{color}

\title[Photometric study of the Blazhko star {MW~Lyr} I.]{An extensive 
photometric study of the Blazhko RR Lyrae star MW~Lyr:
I. Light curve solution}
\author[J. Jurcsik et al.]{J. Jurcsik$^{1}$\thanks{E-mail: jurcsik@konkoly.hu},
\'A. S\'odor$^{1}$,  Zs. Hurta$^{1,2}$,  M. V\'aradi$^{1,3}$, B. Szeidl$^{1}$, H. A. Smith$^{4}$, \and A. Henden$^{5}$,
I. D\'ek\'any$^{1}$, I. Nagy$^{2}$, K. Posztob\'anyi$^{6}$, A. Szing$^{7}$, K. Vida$^{1,2}$, \and and N. Vityi$^{2}$\\
\\
$^{1}$Konkoly Observatory of the Hungarian Academy of Sciences, H--1525 Budapest PO Box 67, Hungary\\
$^{2}$Dept. of Astronomy, E\"otv\"os University, , H--1518 Budapest PO Box 49, Hungary\\
$^{3}$Observatoire de Geneve, Universite de Gen\`eve, CH--1290, Sauverny, Switzerland\\
$^{4}$Dept. of Physics and Astronomy, Michigan State Univ., East Lansing, MI 48824, USA\\
$^{5}$American Association of Variable Star Observers, 49 Bay State Road, Cambridge, MA 02138, USA\\
$^{6}$AEKI, KFKI Atomic Energy Research Institute, Thermohydraulic Department, H--1525 Budapest 114, PO Box 49, Hungary\\
$^{7}$University of Szeged, Dept. of Exp. Physics and Astron. Obs., H--6720 Szeged, D\'om t\'er 9, Hungary
}
\begin{document}

\date{Accepted 2008 ..... Received 2008 ...; in original form 2008 May 15}

\pagerange{\pageref{firstpage}--\pageref{lastpage}} \pubyear{2008}

\maketitle

\label{firstpage}

\begin{abstract}
We have obtained the most extensive and most accurate photometric data of a Blazhko variable
MW Lyr during the 2006-2007 observing seasons. The data within each 0.05 phase bin of the 
modulation period ($P_m=f_m^{-1}$) cover the entire light cycle of the primary pulsation period 
($P_0=f_0^{-1}$), making possible a very rigorous and complete analysis. 
The modulation period is found to be 16.5462 d, which is about half of that was reported earlier
from visual observations. Previously unknown 
features of the modulation have been detected. Besides the main modulation frequency 
$f_\mathrm{m}$,  sidelobe modulation frequencies around the pulsation frequency and its 
harmonics appear at $\pm2f_\mathrm{m}$, $\pm4f_\mathrm{m}$, and $\pm12.5f_\mathrm{m}$ 
separations as well. Residual signals in the prewhitened light curve larger than the 
observational noise appear at the minimum-rising branch-maximum phase of the pulsation, 
which most probably arise from some stochastic/chaotic behaviour of the pulsation/modulation. 
The Fourier parameters of the mean light curve differ significantly from the averages of
the Fourier parameters of the observed light curves in the different phases of the Blazhko 
cycle. Consequently, the mean light curve of MW Lyrae never matches its actual 
light variation. The $\Phi_{21}$, $\Phi_{31}$ phase differences in different phases of 
the modulation show unexpected stability during the Blazhko cycle. A new phenomenological 
description of the light curve variation is defined that separates the amplitude and phase 
(period) modulations utilising the phase coherency of the lower order Fourier phases. 
\end{abstract}


\section{Introduction}

The Blazhko modulation of RR Lyrae stars, a phenomenon known for about a century, is still 
one of the open questions in astrophysics. The problems and the inconsistencies of the existing 
models with observational facts were recently discussed in details by \cite{stothers}, 
therefore we only briefly mention here that none of the suggested models
(magnetic oblique rotator, \citet{shiba}; resonant excitation of nonradial modes, \citet{dz})
can explain the complexity of all the observed properties of the modulation.

Alternatively,  \cite{stothers} suggests that the modulation may be explained by continuous
amplitude and period changes of the pulsation due to the action of a turbulent convective dynamo
in the lower envelope of the star, with dynamo cycle identical with the Blazhko period.
This explanation does not involve any nonradial mode component, the phenomenon is interpreted 
in the framework that Blazhko RRab stars are purely fundamental mode radial pulsators.
However, \cite{stothers} gives only a qualitative, rough picture of his model, that have to be 
checked both observationally and theoretically, thus the investigation of the phenomenon still 
remains  an important and valid task.

A lot of efforts have been already made in studying the light curve changes of RR Lyrae stars 
but most of the available photometries of Blazhko variables have some defects: inaccuracy 
(visual, photographic data), biased data sampling (the photoelectric observations focused mostly 
on the rising branch, maximum phase of the light curve), data inhomogeneity due to rare data 
sampling on a too long timebase (the time scale of period changes and changes in the modulation 
properties can be as short as a few years), etc. Most of the recent CCD and/or photoelectric 
observations of individual Blazhko stars are not extended enough to study the modulation 
properties in full detail. The first multicolour photometric data that covered each phase of 
both the modulation and the pulsation within a short time-base (one season) were published for 
RR Gem \citep{rrgI}. During the period of the CCD observations RR Gem showed, however, small 
amplitude modulation, that limits the available information due to the low S/N ratios of the 
modulation signals.

We started a systematic search for previously unknown Blazhko variables at Konkoly 
Observatory in 2004 using the advantage of  full access to an automated 60 cm telescope 
\citep{sodor}. The aim has also been to clear up questionable cases.
In \citet{ibvs} we have revised the list of known Blazhko variables \citep{smith}.  
The modulation of MW~Lyrae was found to be ambiguous as the photographic observations 
\citep{gessner} did not confirm the modulation detected in visual data by \cite{mandel}. 
Our CCD observations of MW Lyrae affirm the light curve modulation of the star, but with a 
modulation period about half of that \cite{mandel} announced. 

Spectroscopic observation of MW Lyr has never been obtained, however, based on its short 
period ($P_{\mathrm puls}=0.397$~days) it should be a relatively hot and metal rich RR Lyrae star. 

In the 2006 and 2007 seasons we obtained CCD photometric observations of MW~Lyrae 
\hbox{($\alpha=18^h 19^m 53\fs8, \delta=+31\degr 58\arcmin 54\arcsec$, J2000)}.
This is the first multicolour photometric data set of a large modulation amplitude Blazhko 
variable that is  condensed, extended, and accurate enough to detect previously unknown 
properties of the modulation.

In this  paper  the photometric data of MW Lyr are published and analysed. The light curve 
solution (frequency analysis) is discussed using mostly the $V$ band data. A second paper is 
going to be devoted to the study of the colour behaviour during the course of the modulation cycle. 

\begin{table*}
  \caption{Data of comparison and check stars}
 \label{comps}
  \begin{tabular}{llclcccc}
  \hline
   Name &GSC2.2 ID &USNO-B1.0 ID   & \multicolumn{1}{c}{  coord } & $V$[mag]$^*$&$B-V$[mag]$^*$& $V-R_{\rm C}$[mag]$^*$&$V-I_{\rm C}$[mag]$^*$\\
  &&&      RA \, \, \, \, \, \, \, \, \, Dec     & &&&\\
 \hline
C0&N0223233663 &1219-0328347 &18 19 56.80 \,\, +31 58 13.3&$11\fm841$&$0\fm582$&$0\fm315$&$0\fm666$\\
C1&N0223233659 &1219-0328429 &18 20 06.56 \,\, +31 58 56.4&$12\fm986$&$0\fm682$&$0\fm383$&$0\fm788$\\
C2&N0223233671 &1219-0328258 &18 19 48.61 \,\, +31 57 14.6&$13\fm921$&$0\fm858$&$0\fm480$&$0\fm950$\\
C3&N02232332613&1219-0328267 &18 19 48.91 \,\, +31 56 54.4&$14\fm363$&$0\fm603$&$0\fm338$&$0\fm671$\\
\hline
\multicolumn{8}{l}{\scriptsize $^*$ Standard magnitudes measured by A. Henden}\\

\end{tabular}
\end{table*}

\begin{figure}
  \includegraphics[width=7.5cm]{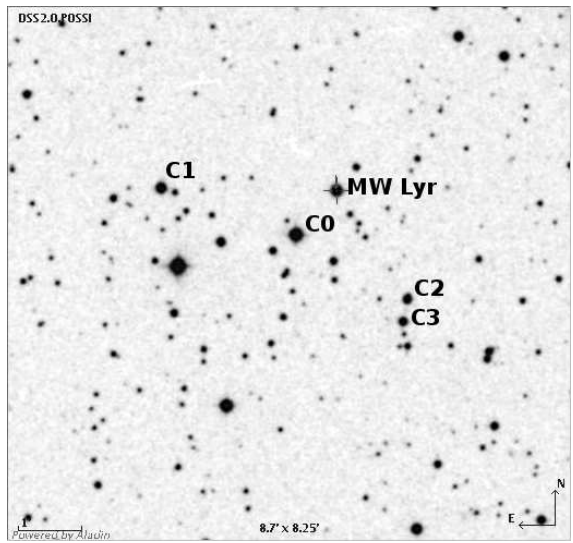}
\center
  \caption{Map of MW~Lyr and the comparison (C0) and check stars (C1,C2, and C3). }
\label{chart}
\end{figure}

\section{Data}
The major part of the observations were obtained with the automated 60 cm telescope of the Konkoly 
Observatory, Sv\'abhegy, Budapest equipped with a Wright Instruments $750\times1100$ CCD camera 
(FoV $17\arcmin\times24\arcmin$). Measurements were taken on 177 nights between May 2006 and Jul 2007.
CCD observations with the 60 cm telescope of the Michigan State University equipped with an 
Apogee Ap47p CCD camera (FoV $10\arcmin\times10\arcmin$) and with the 1m RCC telescope of the Konkoly 
Observatory equipped with a Princeton Instruments VersArray 1300B CCD camera (FoV $6\arcmin\times6\arcmin$) 
were obtained on 6 and 3 additional nights in July 2006 and in Aug 2007, respectively. Johnson-Cousins 
$BVI_\mathrm{C}$ filters were used in all the observations. Altogether 5700-5800 data points in the 
$BVI_{\rm C}$ passbands were gathered. Observations in $R_{\rm C}$ band were also obtained on 15/3 nights 
with the Konkoly 60/100 cm telescopes.

Data reduction was performed using standard IRAF\footnote{{\sc IRAF} is distributed by the National 
Optical Astronomy Observatory, which is operated by the Association of Universities for Research 
in Astronomy, Inc., under cooperative agreement with the National Science Foundation.} packages. 
Aperture photometry of MW~Lyr and several neighbouring stars were carried out. In the analysis 
magnitudes of MW~Lyr relative to  GSC2.2 N0223233663 (C0) are used. Table~\ref{comps} lists the basic data and 
Fig.~\ref{chart} shows the positions of our comparison and check stars. The magnitude differences of 
the comparison and check stars remained constant within 0.010-0.015, 0.009-0.013, 0.008-0.012, 
0.009-0.013 mag  in the $B, V, R_C$, and $I_\mathrm{C}$  bands, respectively (see also Fig.~\ref{residualoh}).

Transformation to the standard system was done using the $B, V, R_{\rm C}$, and $I_{\rm C}$ magnitudes  
of the stars in the field of MW Lyr observed by A. Henden with the USNO Flagstaff Station 1.0 m 
telescope equipped with a SITe/Tektronix  $1024\times1024$ CCD. A complete list of the positions 
and standard $BV(RI)_{\rm C}$ magnitudes of these stars are available online as Supplementary Material (Table 1a). 
To calculate the colour terms of the transformations the nightly instrumental $V$ and $B$ light curves 
were fitted by different order Fourier sums to determine the $V$ magnitudes at the moments of the 
$B, R_{\mathrm C}, I_{\mathrm C}$ observations and the $B$ magnitudes at the moments of the $V$ 
observations. No transformation was applied in the $I_{\rm C}$ band of the Konkoly 60 cm data as no 
colour dependency of the differences between the instrumental and standard magnitudes was found 
in this band.

Taking into account the proximity of the comparison star to the variable only second order extinction 
correction were applied in the B band. 

$B-V$, $V-R_{\rm C}$, and $V-I_{\rm C}$ colours were  derived utilizing the $V$ observations and fitted 
values of the $B$, $R_{\rm C}$, $I_{\rm C}$ curves according to nightly Fourier fits for the moments of 
the $V$ measurements. Differential $V$ magnitudes and $B-V$, $V-R_{\rm C}$, and $V-I_{\rm C}$ 
colours of MW~Lyr with respect to C0 are given in Table~\ref{phot}. The Heliocentric Julian Date,  $\Delta V$, 
$\Delta (B-V)$, $\Delta (V-R_{\rm C})$, and $\Delta (V-I_{\rm C})$ relative magnitudes, and the observatory ID 
are given in Column 1-6, respectively. The entire list of photometric data is available online as 
Supplementary Material. In Table 2a, 2b, and 2c (electronic only)  the $\Delta B$, 
$\Delta R_{\rm C} $ and $\Delta I_{\rm C} $ time series are given.

\begin{table}
\caption{Time series of the $V$ magnitude and colour differences of MW Lyrae relative to the comparison star C0} 
 \label{phot}
\begin{minipage}{100mm}
  \begin{tabular}{l@{\hspace{7pt}}c@{\hspace{6pt}}c@{\hspace{6pt}}c@{\hspace{6pt}}c@{\hspace{5pt}}c}
  \hline
HJD-2\,400\,000& $\Delta V$& $\Delta (B-V)$& $\Delta (V-R_\mathrm{C})$ &$\Delta (V-I_\mathrm{C})$&obs$^{*}$\\
 \hline
53887.34009  &  2.218  & -0.075 & -0.064 & -0.041  &1\\
53887.35249  &  2.227  & -0.073 & -0.035 & -0.035  &1\\
53887.35867  &  2.250  & -0.087 & -0.020 & -0.020  &1\\
53887.36489  &  2.257  & -0.085 & -0.020 & -0.023  &1\\
...&...&...&...&...&....\\
\hline
\multicolumn{5}{l}{\scriptsize $^*$ 1) Konkoly 60cm; 2) MSU 60cm; 3) Konkoly 1m}\\
\end{tabular}
\end{minipage}
\end{table}

Maximum timings and maximum brightness values of the $V$  data set were determined for 88 epochs (Table~\ref{maxtb}).
The complete list of these data is available online.
\begin{table}
  \caption{Maximum timings and brightness values derived from the $V$ light curve.}
 \label{maxtb}
  \begin{tabular}{cc}
  \hline
maximum time& $V$ maximum brightness\\
HJD-2\,400\,000&  relative magnitude to C0\\
 \hline
53887.520 &1.501\\
53901.447 &1.285\\
53903.428 &1.480\\
...&...\\
\hline
\end{tabular}
\end{table}

\begin{figure}
  \includegraphics[width=9cm]{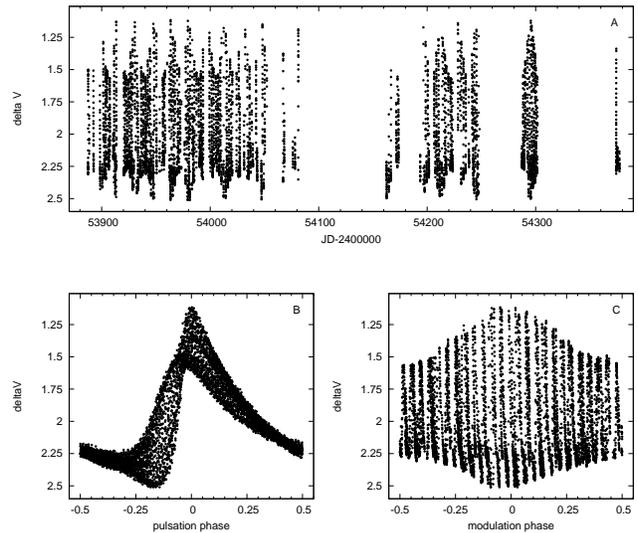}
  \caption{Delta $V$ magnitudes versus Julian Date (panel A), and data phased with the 0.397674 d
pulsation (panel B), and the \hbox{16.546\,d} modulation periods (panel C) are shown. }
\label{lc}
\end{figure}

\begin{figure}
  \includegraphics[width=6.8cm]{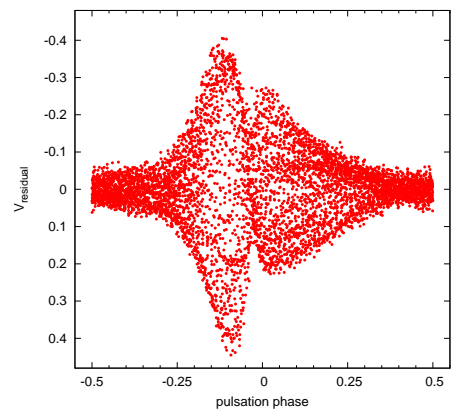}
  \caption{Residual $V$ light curve  of MW~Lyr after removing the pulsation components of the light 
curve solution from the data. The residuals have larger amplitude at around the minimum, rising branch 
phases of the pulsation than around maximum brightness due to the large amplitude of the phase modulation. }
\label{lepke}
\end{figure}

\section{Results}

\begin{table*}
\caption{Light curve solutions}
 \label{solutions}
  \begin{tabular}{l@{}cl}
  \hline
Solution& No. of frequencies & Frequency identification\\
 \hline
A (triplet) & 36 & $kf_0, k=1,...12; kf_0-f_\mathrm{m}, k=1,...10; kf_0+f_\mathrm{m}, k=1,...13; f_\mathrm{m}$\\
B (quintuplet) & 53 &  frequencies in solution A and $kf_0-2f_\mathrm{m}, k=1,...7; kf_0+2f_\mathrm{m}, k=1,...9; 2f_\mathrm{m}$\\
C (septuplet) & 58 & frequencies in solution B and  $kf_0-4f_\mathrm{m}, k=1,2,3; kf_0+4f_\mathrm{m}, k=1,2$\\
D  & 66 & frequencies in solution C and $kf_0-12.5f_\mathrm{m}, k=2,3; kf_0+12.5f_\mathrm{m}, k=1,2; kf_0-f_\mathrm{m}', k=1,2,3; f_\mathrm{m}'$\\
E  & 96 &  frequencies in solution D and 30 additional frequencies (see details in the text)\\
\hline
\multicolumn{3}{l}{$f_0=2.514621$\,cd$^{-1}$;\,\, $f_\mathrm{m}=0.060437$\,cd$^{-1}$;\,\, $f_\mathrm{m}'= 0.00197$\,cd$^{-1}$} \\
\hline
\end{tabular}
\end{table*}

The $V$ light curve of MW~Lyrae and the data folded with the pulsation and modulation periods are shown 
in Fig.~\ref{lc}. The light curve is strongly modulated, the full amplitudes of the amplitude and phase 
modulations are larger than 0.45 mag and 0.07 phase of the pulsation ($\sim$ 40 min), respectively (see 
also Fig.~\ref{egg}). The modulation seems to have larger amplitude at around minimum and rising branch 
phases than at maximum light as Fig.~\ref{lepke} shows. In this figure the residual light curve is phased 
with the pulsation period after the pulsation components are removed. The large amplitude of the residuals 
around minimum phase is the consequence of the large amplitude of the phase modulation. In reality, the 
amplitude of the maximum brightness variation is larger than the amplitude of the minimum brightness 
variation as the top and bottom envelope curves of panel C in Fig.~\ref{lc} show.

The elements of the pulsation and modulation are: 

\begin{displaymath}
T_{\mathrm{max\,puls}} = 2\,453\,963.4950\,{\mathrm [HJD]} + 0.3976742\cdot E_{\mathrm{puls}},
\end{displaymath}

and

\begin{displaymath}
T_{\mathrm{max\,Bl}} = 2\,453\,963.4950\,{\mathrm [HJD]} + 16.5462 \cdot E_{\mathrm{Bl}}.
\end{displaymath}

The pulsation and modulation periods are those that yield the best fit to the $V$ light curve with the 
pulsation ($kf_0$) and modulation ($kf_0\pm f_\mathrm{m}$,  $kf_0\pm 2f_\mathrm{m}$, $f_\mathrm{m}$, and 
$2f_\mathrm{m}$)  frequency components using `locked' frequency solution (i.e., the modulation components 
are at the positions of the linear combination frequencies). Details of the frequency component determinations 
are given in Sect 3.1.

Data analysis was performed using the different applications of the MUFRAN/TIFRAN packages \citep{mufran,tifran}, 
a linear combination fitting program developed by \'A. S\'odor, and the linear and nonlinear curve fitting 
abilities of gnuplot\footnote{\tt http://www.gnuplot.info/}.

\subsection{The light curve solution}

\begin{figure*}
  \centering
  \includegraphics[width=19.0cm]{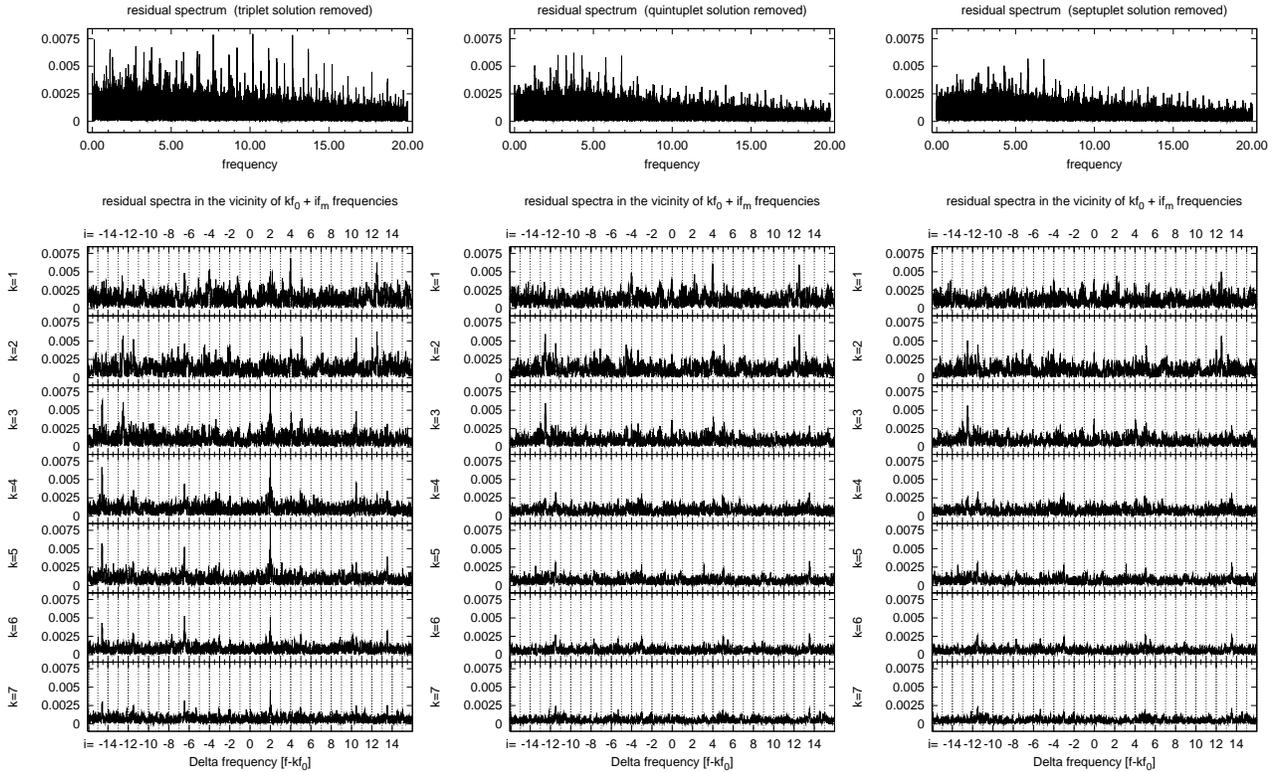}
  \caption{Residual spectra (top panels) and spectra in the vicinity of  the $kf_0$, $k=1,...7$ pulsation 
frequency components (bottom panels) are drawn. In these panels $\sim 2$ cd$^{-1}$ frequency ranges are 
enlarged. On the X axis ticks are at $2f_\mathrm{m}$ separation, 0 corresponds to the positions of $kf_0$.
For the sake of lucidity a grid with $if_\mathrm{m}$ spacing is also drawn.  In the left panels residuals 
after prewhitening with the triplet frequency solution (solution A) are shown. The highest peaks in the 
spectrum appear at $kf_0+2f_\mathrm{m}$ frequencies. The  $kf_0-2f_\mathrm{m}$ frequencies do not evidently 
show up in these spectra but in later steps of the prewhitening some of these components also emerge with 
S/N ratio larger than 3. Residuals after prewhitening with the quintuplet frequency solution (solution B) 
are shown in the middle panels. The highest peaks appear at $kf_0\pm4f_\mathrm{m}$ and $kf_0\pm12.5f_\mathrm{m}$ 
frequencies. These two series of modulation components are unfortunately seriously biased as they are very 
close to the $\pm$1 cd$^{-1}$ alias components of each other. However, after prewhitening the data also  
with 5 frequencies of the $kf_0\pm4f_\mathrm{m}$ components (septuplet: solution C) as shown in the 
right panels, it becomes evident that the  $kf_0\pm12.5f_\mathrm{m}$  components still remain in 
the spectrum. We conclude therefore, that they are real frequencies, indeed, and not alias artifacts of the 
data sampling. There are other well defined frequency series appearing with common frequency separations, too. 
One is very close to the pulsation components at $f_\mathrm{m}'=0.0019$ cd$^{-1}$ separation.
Four components of this series can be identified:  $f_0-f_\mathrm{m}'$, $2f_0-f_\mathrm{m}'$, 
$3f_0-f_\mathrm{m}'$ and $f_\mathrm{m}'$.  In light curve solution D the septuplet and the  
$kf_0\pm12.5f_\mathrm{m}$ modulation and also modulation components with  $f_\mathrm{m}'=0.00197$\,cd$^{-1}$ 
separation are removed. The residuals of light curve solution D are shown in Fig.~\ref{residualszin}.}
  \label{residual}
\end{figure*}

\begin{figure*}
  \centering
  \includegraphics[width=19.0cm]{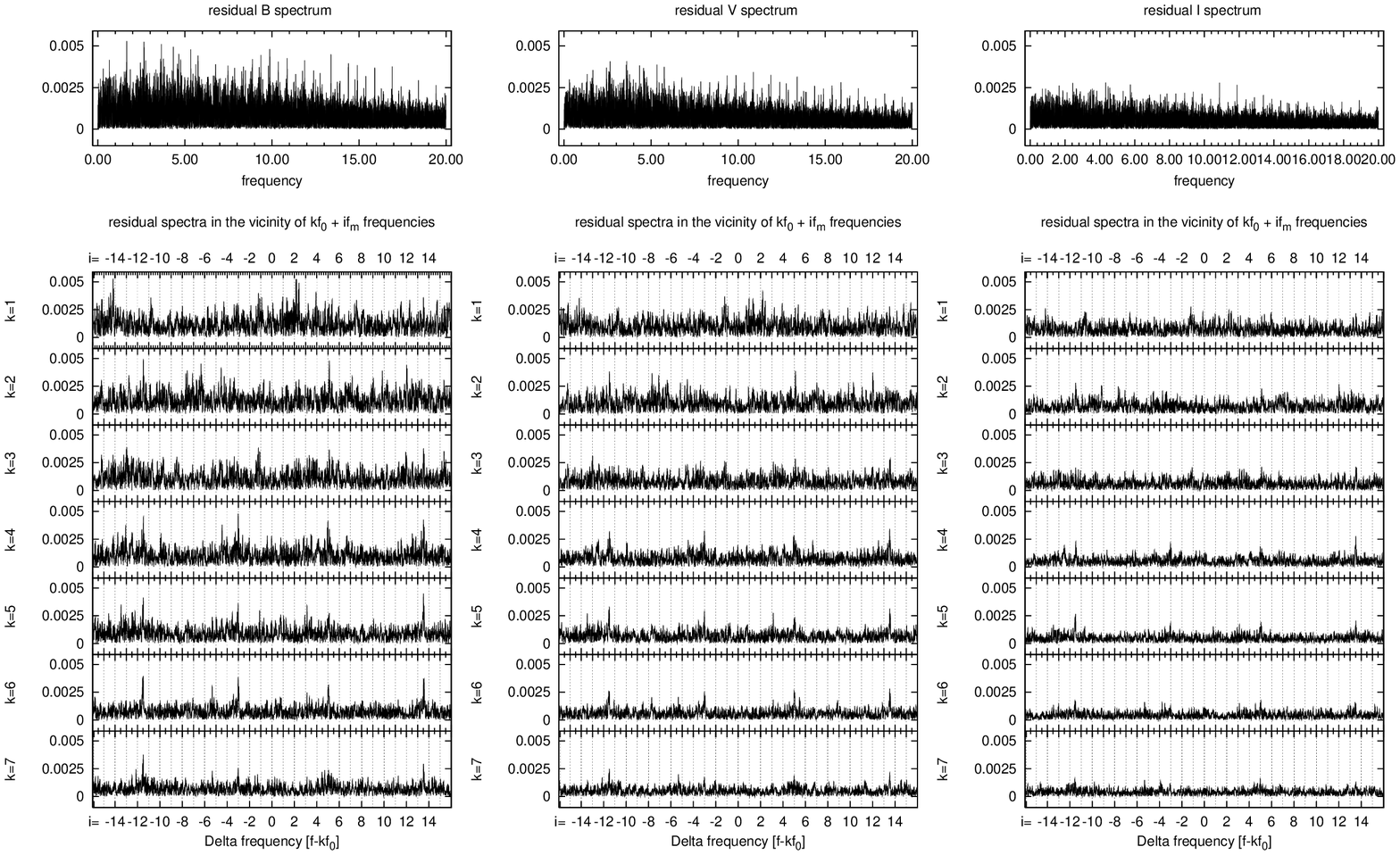}
 \caption{The top panels show the residual spectra of the $B$, $V$ and $I_\mathrm{C}$ data prewhitened with 
frequency solution D (septuplet frequencies and modulation components with $12.5f_\mathrm{m}$ and 
$f_\mathrm{m}'$ separations). The spectra are zoomed in the vicinity of the $kf_0$ frequencies $k=1,...7$ 
in the bottom panels. Not regarding the decreasing mean level of the spectra towards longer wavelengths 
the residuals are similar, peaks at $-3f_\mathrm{m}$, $5f_\mathrm{m}$, $-11.5f_\mathrm{m}$, and $13.5f_\mathrm{m}$ 
can be identified. Further significant peaks appear in the  $|\Delta f|<3f_\mathrm{m}$  vicinity of $f_0$.
Though it is very probable that many of these frequencies are real signals we stop with frequency 
identifications here as the addition of these components does not significantly improve the light curve solution.}
  \label{residualszin}
\end{figure*}

The Fourier spectrum of Blazhko RRab stars is characterized by equidistant frequency triplets 
with frequency separation identical with the modulation frequency \citep[see e.g.][]{arh,rrgI,rrl}. 
In the residual spectrum of  RV~UMa, equidistant quintuplets were identified by \cite{rvuma}. 
The appearance of the modulation frequency ($f_\mathrm{m}$) itself was a matter of debate for long, but in the 
spectra of extended and accurate datasets  $f_\mathrm{m}$ also shows up unquestionably \citep{rrgI,sscnc,rvuma}.

The Fourier spectrum of MW~Lyrae is also dominated by  these frequency components. Pulsation frequencies appear up 
to the 12th order, while the $kf_0+f_\mathrm{m}$ and $kf_0-f_\mathrm{m}$ components are present with $k\le 13$ and 
$k\le 10$, respectively. The modulation frequency ($f_\mathrm{m}$) has an amplitude of 0.014 mag in the $V$ data, 
which is as high as the amplitude of the 6th harmonic component of the pulsation. Prewhitening the data with the 
frequencies of the triplet solution significant peaks in the residual spectrum appear, indicating that the light 
curve cannot be accurately fitted simply with equidistant frequency triplets.

Fig.~\ref{residual} shows the residual spectra of the $V$ data in the $0 - 20$ cd$^{-1}$ frequency range and in the 
vicinity of the $kf_0$ ($k=1,...7$) pulsation frequencies after prewhitening with different frequency solutions. 
Table~\ref{solutions} summarizes the frequency solutions applied as more and more frequency components are identified. 
All the identified frequency components (pulsation plus modulation frequencies) are simultaneously fitted to the
original data and this light curve solution is used in the next step of the analysis. In Sect 3.3 it is documented 
that the simultaneous fit of all the frequency components gives better light curve solution than if the frequency 
components are successively fitted and are removed from the data in consecutive steps.
 
The left panels of Fig.~\ref{residual} show the residual spectrum after removing the triplet frequency solution 
(solution A). The highest peaks appear at $kf_0+2f_\mathrm{m}$ frequencies. The $kf_0-2f_\mathrm{m}$ and  
the $2f_\mathrm{m}$ frequencies are also detected  but with relatively small amplitudes. There are altogether 
16 frequencies identified to belong to the $kf_0\pm2f_\mathrm{m}$ modulation series. Fitting the data with the 
quintuplet frequencies (solution B) the residual spectrum (middle panels in Fig.~\ref{residual}) is still not flat, 
besides other peaks two sets of frequencies are evident, one at $4f_\mathrm{m}$ and the other at $12.5f_\mathrm{m}$ 
frequency separations. Unfortunately, the frequencies of these series are very close to the $\pm1$ cd$^{-1}$  alias 
components of each other (see their frequency values in Table~\ref{big}), that makes the determination of the 
amplitudes of these components ambiguous. However, either the $kf_0\pm4f_\mathrm{m}$ or the 
$kf_0\pm12.5f_\mathrm{m}$ components are removed,  members of the other frequency series remain, consequently the 
components of  the  $kf_0\pm4f_\mathrm{m}$ and the $kf_0\pm12.5f_\mathrm{m}$ frequency series are independent 
signals. In the right panels in Fig.~\ref{residual} it is shown that even if the septuplet solution is removed  
with 5 components of the $kf_0\pm4f_\mathrm{m}$ series  (solution C), signals  at $12.5f_\mathrm{m}$ separations 
are still present in the residual. These are the highest peaks detected in this spectrum. It is not at all clear 
whether these modulation components are indeed connected to the main modulation frequency or it happens just by 
chance that this secondary modulation has a modulation frequency very close to $12.5 f_\mathrm{m}$.
We can only say that the frequencies of this modulation series are within the uncertainties at the positions of  
$kf_0\pm12.5f_\mathrm{m}$. If the frequencies of these modulation components are not locked to the  
$kf_0\pm12.5f_\mathrm{m}$ positions in the fitting process their displacements do not exceed significantly and/or 
systematically the displacements of the other modulation frequency components with similar amplitude 
(see data in Table~\ref{big}).

Some of the other peaks in the residual of the septuplet solution also form series with common frequency 
separation from the pulsation components. There are remaining peaks very close to the pulsation components at 
$f_0-f_\mathrm{m}'$, $2f_0-f_\mathrm{m}'$, $3f_0-f_\mathrm{m}'$, and  $f_\mathrm{m}'$ frequencies with 
$f_\mathrm{m}'=0.00197$ cd$^{-1}$. The periods and amplitudes of these frequency components are somewhat uncertain 
as this modulation period is $\sim500$~days, hardly shorter than the total length of the observations.
These components can be identified either with an additional long period modulation or with residuals caused 
by slight changes of the pulsation period during the observations. 

In light curve solution D, besides the quintuplet frequencies, the $kf_0\pm12.5f_\mathrm{m}$ and the 
$f_\mathrm{m}'$ modulation components are also fitted and removed. In Fig.~\ref{residualszin} the residual spectra 
of the $B$, $V$, and $I_\mathrm{C}$ observations are shown after prewhitening with light curve solution D.
These residual spectra are characterized with a broad band low frequency signal centred on about the 
pulsation frequency and with series of further modulations at e.g.,  $-3f_\mathrm{m}$, $5f_\mathrm{m}$, 
$-11.5f_\mathrm{m}$, and $13.5f_\mathrm{m}$ frequencies. There are also further frequency peaks in the 
$\pm0.15$ cd$^{-1}$ vicinity of the $f_0$ pulsation frequency. Though the S/N ratio of the amplitudes of 
many of these frequency components are larger than 3, we stop frequency identification at this level 
as the addition of these frequency components does not improve the light curve solution significantly.

In Table~\ref{big} details of light curve solution D are given for the $B$, $V$, and $I_\mathrm{C}$ data. 
Col.~1 gives the identification of the frequencies, Col.~2 lists the frequencies of the `locked' frequency 
solution (exact linear combinations of the pulsation and modulation components). The next 3 columns give  
the differences in the frequency  values  ($\Delta f$) of the `let free' solutions for the $B$, $V$, and 
$I_\mathrm{C}$ light curves. In the let free solutions only the harmonic components of the pulsation are at 
locked frequency values but the best frequency values of $f_0$ and all the modulation frequency components 
are searched in a nonlinear process. For comparison purposes, $\Delta f/\sigma_f$ are given in Cols 6-8 
in the three bands. The $\sigma_f$ error estimates of the frequencies are calculated using the formula 
given by \cite{md}. It is important to note here that, in the case of correlated noise, these errors 
underestimate the true uncertainties of the frequencies significantly. For most of the identified modulation 
frequency components  the frequency displacements are 0.5 - 4.0 times the calculated $\sigma_f$ values.

Frequency components with larger displacements ($\Delta f/\sigma_{f}>3$) in each band are:\\
$f_\mathrm{m}$ ($V_{amp}=0.014$); $\Delta f(V)=0.00011$ cd$^{-1}$\\
$f_0+2f_\mathrm{m}$ ($V_{amp}=0.004$); $\Delta f(V)=-0.00055$ cd$^{-1}$\\ 
$4f_0-2f_\mathrm{m}$ ($V_{amp}=0.003$); $\Delta f(V)=-0.00064$ cd$^{-1}$\\
$f_0-4f_\mathrm{m}$ ($V_{amp}=0.003$); $\Delta f(V)=0.00041$ cd$^{-1}$\\ 
$3f_0-4f_\mathrm{m}$ ($V_{amp}=0.002$); $\Delta f(V)=0.00090$ cd$^{-1}$ \\
$f_\mathrm{m}'$ ($V_{amp}=0.002$); $\Delta f(V)=0.00196$ cd$^{-1}$.\\

The larger uncertainty of the $f_\mathrm{m}'$ component arises simply from the fact that this modulation period 
is hardly shorter than the time span of the observations. The larger displacements of the $4f_\mathrm{m}$  
components are most probably due to their strong $\pm 1$ cd$^{-1}$ alias connections with the frequencies of 
the $12.5f_\mathrm{m}$ modulation series. The reason why the frequency displacement of the other three 
modulation components are unexpectedly large is unknown. Based on the data given in Table~\ref{big}, we think that 
there is no serious reason to assume that the modulation frequencies are not, in fact, at their `locked' positions.

\begin{table*}
\scriptsize
 \begin{minipage}{220mm}
  \caption{Fourier parameters of light curve solution D in the $B,V,I_{\mathrm C}$ bands}
 \label{big}
  \begin{tabular}{rrrrrrrrcrcrcr}
  \hline
Frequency ID & Frequency &$\Delta f(B)$&$\Delta f(V)$&$\Delta f(I_C)$&$\Delta f/\sigma_{f}$ &$\Delta f/\sigma_{f}$ &$\Delta f/\sigma_{f}$ & $A(B)$ & $\Phi (B)$& $A(V)$ & $\Phi (V)$ & $A(I_C)$& $\Phi (I_C)$ \\
&[cd$^{-1}$]&\multicolumn{3}{c}{[$10^{-5}$ cd$^{-1}$]}&$B$&$V$&$I_\mathrm{C}$&[mag]&[rad]&[mag]&[rad]&[mag]&[rad]\\
\hline
$f_0$ & $2.514621$ & $0.14$ & $0.39$ & $0.56$ & $2.0$ & $4.2$ & $3.7$ & $0.4998$ & $1.963$ & $0.3741$ & $1.919$ & $0.2302$ & $1.772$ \\
$2f_0$ & $5.029242$ & -- & -- & -- & -- & -- & -- & $0.2063$ & $6.211$ & $0.1585$ & $6.200$ & $0.1006$ & $6.152$ \\
$3f_0$ & $7.543863$ & -- & -- & -- & -- & -- & -- & $0.0968$ & $4.366$ & $0.0759$ & $4.354$ & $0.0500$ & $4.335$ \\
$4f_0$ & $10.058484$ & -- & -- & -- & -- & -- & -- & $0.0469$ & $2.323$ & $0.0373$ & $2.334$ & $0.0251$ & $2.317$ \\
$5f_0$ & $12.573105$ & -- & -- & -- & -- & -- & -- & $0.0242$ & $0.305$ & $0.0199$ & $0.268$ & $0.0138$ & $0.268$ \\
$6f_0$ & $15.087726$ & -- & -- & -- & -- & -- & -- & $0.0165$ & $4.446$ & $0.0139$ & $4.388$ & $0.0093$ & $4.389$ \\
$7f_0$ & $17.602347$ & -- & -- & -- & -- & -- & -- & $0.0117$ & $2.360$ & $0.0096$ & $2.394$ & $0.0069$ & $2.332$ \\
$8f_0$ & $20.116968$ & -- & -- & -- & -- & -- & -- & $0.0093$ & $0.324$ & $0.0074$ & $0.294$ & $0.0049$ & $0.368$ \\
$9f_0$ & $22.631589$ & -- & -- & -- & -- & -- & -- & $0.0065$ & $4.670$ & $0.0055$ & $4.696$ & $0.0039$ & $4.574$ \\
$10f_0$ & $25.146210$ & -- & -- & -- & -- & -- & -- & $0.0053$ & $2.735$ & $0.0039$ & $2.856$ & $0.0025$ & $2.592$ \\
$11f_0$ & $27.660831$ & -- & -- & -- & -- & -- & -- & $0.0031$ & $0.974$ & $0.0028$ & $0.788$ & $0.0022$ & $0.940$ \\
$12f_0$ & $30.175452$ & -- & -- & -- & -- & -- & -- & $0.0024$ & $5.215$ & $0.0018$ & $5.329$ & $0.0010$ & $5.384$ \\
$f_\mathrm{m}$ & $0.060437$ & $8.87$ & $10.95$ & $14.93$ & $4.9$ & $4.4$ & $4.3$ & $0.0193$ & $3.628$ & $0.0140$ & $3.635$ & $0.0100$ & $3.558$ \\
$f_0 + f_\mathrm{m}$ & $2.575058$ & $0.06$ & $0.26$ & $-0.04$ & $0.2$ & $0.7$ & $-0.1$ & $0.1217$ & $3.316$ & $0.0900$ & $3.332$ & $0.0559$ & $3.352$ \\
$2f_0 + f_\mathrm{m}$ & $5.089679$ & $0.83$ & $0.27$ & $0.24$ & $2.6$ & $0.6$ & $0.3$ & $0.1067$ & $1.402$ & $0.0805$ & $1.436$ & $0.0506$ & $1.501$ \\
$3f_0 + f_\mathrm{m}$ & $7.604300$ & $0.73$ & $0.57$ & $0.83$ & $1.4$ & $0.8$ & $0.8$ & $0.0648$ & $6.013$ & $0.0502$ & $6.036$ & $0.0324$ & $6.103$ \\
$4f_0 + f_\mathrm{m}$ & $10.118921$ & $3.17$ & $1.87$ & $2.58$ & $4.1$ & $1.9$ & $1.7$ & $0.0451$ & $4.316$ & $0.0357$ & $4.329$ & $0.0231$ & $4.352$ \\
$5f_0 + f_\mathrm{m}$ & $12.633542$ & $2.27$ & $2.94$ & $4.63$ & $1.8$ & $1.9$ & $1.9$ & $0.0274$ & $2.502$ & $0.0219$ & $2.497$ & $0.0142$ & $2.491$ \\
$6f_0 + f_\mathrm{m}$ & $15.148163$ & $9.30$ & $6.38$ & $5.81$ & $4.4$ & $2.3$ & $1.4$ & $0.0165$ & $0.533$ & $0.0126$ & $0.513$ & $0.0085$ & $0.525$ \\
$7f_0 + f_\mathrm{m}$ & $17.662784$ & $6.96$ & $4.21$ & $12.06$ & $2.0$ & $1.0$ & $1.9$ & $0.0098$ & $4.765$ & $0.0085$ & $4.762$ & $0.0054$ & $4.763$ \\
$8f_0 + f_\mathrm{m}$ & $20.177405$ & $11.58$ & $5.22$ & $5.85$ & $2.3$ & $0.8$ & $0.6$ & $0.0069$ & $2.584$ & $0.0054$ & $2.627$ & $0.0033$ & $2.676$ \\
$9f_0 + f_\mathrm{m}$ & $22.692026$ & $6.93$ & $0.33$ & $-8.80$ & $1.2$ & $0.0$ & $-0.6$ & $0.0058$ & $0.627$ & $0.0042$ & $0.595$ & $0.0024$ & $0.721$ \\
$10f_0 + f_\mathrm{m}$ & $25.206647$ & $-9.99$ & $9.96$ & $13.16$ & $-1.2$ & $1.0$ & $0.8$ & $0.0041$ & $4.919$ & $0.0035$ & $4.798$ & $0.0022$ & $5.023$ \\
$1f_0 + f_\mathrm{m}$ & $27.721268$ & $-13.31$ & $19.24$ & $28.54$ & $-1.2$ & $1.7$ & $1.4$ & $0.0031$ & $2.820$ & $0.0030$ & $2.894$ & $0.0016$ & $2.586$ \\
$12f_0 + f_\mathrm{m}$ & $30.235889$ & $13.97$ & $2.98$ & $65.67$ & $1.1$ & $0.2$ & $1.8$ & $0.0027$ & $0.681$ & $0.0020$ & $0.987$ & $0.0010$ & $0.608$ \\
$13f_0 + f_\mathrm{m}$ & $32.750510$ & $-9.57$ & $11.16$ & $14.54$ & $-0.6$ & $0.4$ & $0.4$ & $0.0020$ & $5.425$ & $0.0014$ & $5.563$ & $0.0010$ & $5.074$ \\
$f_0 - f_\mathrm{m}$ & $2.454184$ & $3.30$ & $3.74$ & $2.69$ & $6.2$ & $5.2$ & $2.3$ & $0.0650$ & $4.770$ & $0.0486$ & $4.787$ & $0.0299$ & $4.835$ \\
$2f_0 - f_\mathrm{m}$ & $4.968805$ & $0.33$ & $2.23$ & $0.52$ & $0.6$ & $2.9$ & $0.4$ & $0.0599$ & $2.731$ & $0.0451$ & $2.775$ & $0.0278$ & $2.864$ \\
$3f_0 - f_\mathrm{m}$ & $7.483426$ & $1.87$ & $3.54$ & $3.48$ & $2.3$ & $3.4$ & $2.0$ & $0.0426$ & $1.094$ & $0.0331$ & $1.109$ & $0.0204$ & $1.161$ \\
$4f_0 - f_\mathrm{m}$ & $9.998047$ & $-1.28$ & $0.38$ & $-1.02$ & $-1.1$ & $0.2$ & $-0.4$ & $0.0287$ & $5.476$ & $0.0225$ & $5.534$ & $0.0144$ & $5.575$ \\
$5f_0 - f_\mathrm{m}$ & $12.512668$ & $4.19$ & $0.38$ & $-1.26$ & $2.2$ & $0.2$ & $-0.3$ & $0.0184$ & $3.804$ & $0.0141$ & $3.763$ & $0.0094$ & $3.801$ \\
$6f_0 - f_\mathrm{m}$ & $15.027289$ & $-4.38$ & $2.83$ & $-6.53$ & $-1.7$ & $0.8$ & $-1.2$ & $0.0132$ & $1.829$ & $0.0094$ & $1.770$ & $0.0066$ & $1.768$ \\
$7f_0 - f_\mathrm{m}$ & $17.541910$ & $6.88$ & $-1.57$ & $3.95$ & $1.6$ & $-0.3$ & $0.5$ & $0.0081$ & $6.253$ & $0.0070$ & $6.239$ & $0.0047$ & $6.097$ \\
$8f_0 - f_\mathrm{m}$ & $20.056531$ & $-19.32$ & $-14.07$ & $-3.38$ & $-2.8$ & $-1.7$ & $-0.3$ & $0.0049$ & $4.436$ & $0.0042$ & $4.422$ & $0.0026$ & $4.392$ \\
$9f_0 - f_\mathrm{m}$ & $22.571152$ & $-31.04$ & $-21.41$ & $-5.76$ & $-2.5$ & $-1.5$ & $-0.3$ & $0.0028$ & $2.505$ & $0.0025$ & $2.473$ & $0.0017$ & $2.480$ \\
$10f_0 - f_\mathrm{m}$ & $25.085773$ & $-20.99$ & $-1.26$ & $17.14$ & $-1.6$ & $-0.1$ & $0.7$ & $0.0026$ & $0.317$ & $0.0021$ & $0.344$ & $0.0015$ & $6.220$ \\
$2f_\mathrm{m}$ & $0.120874$ & $-2.57$ & $9.02$ & $2.30$ & $-0.7$ & $2.0$ & $0.3$ & $0.0098$ & $0.335$ & $0.0075$ & $0.281$ & $0.0046$ & $0.314$ \\
$f_0 + 2f_\mathrm{m}$ & $2.635495$ & $-67.85$ & $-55.18$ & $-39.37$ & $-8.2$ & $-6.5$ & $-3.9$ & $0.0042$ & $3.216$ & $0.0041$ & $3.317$ & $0.0035$ & $3.499$ \\
$2f_0 + 2f_\mathrm{m}$ & $5.150116$ & $-8.37$ & $-7.21$ & $7.53$ & $-1.0$ & $-0.6$ & $0.5$ & $0.0040$ & $2.666$ & $0.0030$ & $2.502$ & $0.0021$ & $2.788$ \\
$3f_0 + 2f_\mathrm{m}$ & $7.664737$ & $9.36$ & $7.87$ & $8.36$ & $2.6$ & $1.7$ & $1.4$ & $0.0098$ & $1.678$ & $0.0077$ & $1.551$ & $0.0057$ & $1.512$ \\
$4f_0 + 2f_\mathrm{m}$ & $10.179358$ & $-15.74$ & $-7.40$ & $-5.63$ & $-4.3$ & $-1.6$ & $-0.7$ & $0.0096$ & $5.573$ & $0.0075$ & $5.564$ & $0.0043$ & $5.694$ \\
$5f_0 + 2f_\mathrm{m}$ & $12.693979$ & $-4.61$ & $2.84$ & $0.86$ & $-1.3$ & $0.6$ & $0.1$ & $0.0095$ & $4.162$ & $0.0074$ & $4.114$ & $0.0049$ & $4.172$ \\
$6f_0 + 2f_\mathrm{m}$ & $15.208600$ & $-23.02$ & $-17.53$ & $2.52$ & $-4.2$ & $-2.4$ & $0.2$ & $0.0063$ & $2.356$ & $0.0048$ & $2.413$ & $0.0033$ & $2.384$ \\
$7f_0 + 2f_\mathrm{m}$ & $17.723221$ & $-7.92$ & $-0.82$ & $15.79$ & $-1.1$ & $-0.1$ & $1.2$ & $0.0048$ & $0.682$ & $0.0042$ & $0.605$ & $0.0025$ & $0.709$ \\
$8f_0 + 2f_\mathrm{m}$ & $20.237842$ & $-16.68$ & $-18.22$ & $-16.96$ & $-1.6$ & $-1.4$ & $-0.7$ & $0.0033$ & $5.142$ & $0.0026$ & $4.945$ & $0.0013$ & $4.876$ \\
$9f_0 + 2f_\mathrm{m}$ & $22.752463$ & $15.03$ & $17.07$ & $18.32$ & $1.0$ & $0.8$ & $0.5$ & $0.0022$ & $3.303$ & $0.0016$ & $3.363$ & $0.0010$ & $2.844$ \\
$f_0 -2f_\mathrm{m}$ & $2.393747$ & $-2.16$ & $2.55$ & $9.01$ & $-0.3$ & $0.3$ & $0.8$ & $0.0054$ & $5.786$ & $0.0044$ & $5.790$ & $0.0030$ & $5.570$ \\
$2f_0 -2f_\mathrm{m}$ & $4.908368$ & $-15.89$ & $-4.45$ & $-17.27$ & $-3.5$ & $-0.7$ & $-1.6$ & $0.0077$ & $5.134$ & $0.0053$ & $5.054$ & $0.0033$ & $5.351$ \\
$^*3f_0 -2f_\mathrm{m}$ & $7.422989$ & $38.74$ & $37.12$ & $78.85$ & $2.9$ & $1.8$ & $2.3$ & $0.0026$ & $3.328$ & $0.0017$ & $3.572$ & $0.0010$ & $3.798$ \\
$4f_0 -2f_\mathrm{m}$ & $9.937610$ & $-67.38$ & $-63.91$ & $-68.81$ & $-7.3$ & $-5.4$ & $-3.4$ & $0.0038$ & $1.956$ & $0.0029$ & $1.977$ & $0.0017$ & $2.160$ \\
$^*5f_0 -2f_\mathrm{m}$ & $12.452231$ & $-49.39$ & $-56.33$ & $-70.03$ & $-3.1$ & $-2.7$ & $-2.7$ & $0.0022$ & $0.680$ & $0.0016$ & $0.445$ & $0.0013$ & $0.793$ \\
$^*6f_0 -2f_\mathrm{m}$ & $14.966852$ & $-112.05$ & $-123.78$ & $-101.45$ & $-4.4$ & $-4.4$ & $-2.2$ & $0.0014$ & $4.630$ & $0.0012$ & $4.592$ & $0.0008$ & $4.714$ \\
$^*7f_0 -2f_\mathrm{m}$ & $17.481473$ & $-88.32$ & $-94.99$ & $-28.84$ & $-3.5$ & $-3.0$ & $-0.7$ & $0.0014$ & $3.053$ & $0.0011$ & $3.028$ & $0.0008$ & $2.882$ \\
$f_0 + 4f_\mathrm{m}$ & $2.756369$ & $8.49$ & $33.32$ & $8.55$ & $1.8$ & $5.3$ & $1.0$ & $0.0073$ & $5.385$ & $0.0055$ & $5.408$ & $0.0040$ & $5.620$ \\
$2f_0 + 4f_\mathrm{m}$ & $5.270990$ & $10.58$ & $32.28$ & $23.18$ & $1.2$ & $2.9$ & $1.5$ & $0.0040$ & $3.604$ & $0.0031$ & $3.664$ & $0.0022$ & $3.654$ \\
$f_0 -4f_\mathrm{m}$ & $2.272873$ & $98.12$ & $41.49$ & $70.76$ & $7.6$ & $3.7$ & $3.6$ & $0.0027$ & $1.948$ & $0.0031$ & $2.402$ & $0.0018$ & $2.443$ \\
$2f_0 -4f_\mathrm{m}$ & $4.787494$ & $50.96$ & $54.76$ & $-3.99$ & $6.0$ & $3.8$ & $-0.2$ & $0.0041$ & $5.921$ & $0.0024$ & $5.884$ & $0.0021$ & $6.080$ \\
$^*3f_0 -4f_\mathrm{m}$ & $7.302115$ & $88.66$ & $90.46$ & $97.93$ & $7.5$ & $5.1$ & $4.0$ & $0.0029$ & $5.100$ & $0.0020$ & $4.930$ & $0.0014$ & $5.341$ \\
$f_0 + 12.5f_\mathrm{m}$ & $3.270084$ & $-45.30$ & $-20.18$ & $-12.27$ & $-3.7$ & $-1.7$ & $-0.8$ & $0.0028$ & $3.050$ & $0.0030$ & $2.960$ & $0.0024$ & $2.956$ \\
$2f_0 - 12.5f_\mathrm{m}$ & $4.273780$ & $9.48$ & $39.58$ & $50.07$ & $1.2$ & $3.7$ & $2.4$ & $0.0044$ & $3.821$ & $0.0033$ & $3.783$ & $0.0017$ & $4.018$ \\
$^*2f_0 + 12.5f_\mathrm{m}$ & $5.784705$ & $-143.83$ & $-36.28$ & $-101.75$ & $-9.4$ & $-2.8$ & $-5.3$ & $0.0023$ & $1.096$ & $0.0026$ & $1.336$ & $0.0018$ & $1.599$ \\
$3f_0 - 12.5f_\mathrm{m}$ & $6.788401$ & $-17.73$ & $-74.47$ & $-28.12$ & $-2.6$ & $-9.0$ & $-2.1$ & $0.0051$ & $2.557$ & $0.0042$ & $2.473$ & $0.0025$ & $2.346$ \\
$f_0 -f_\mathrm{m}'$ & $2.512651$ & $30.64$ & $39.23$ & $25.88$ & $7.1$ & $5.2$ & $2.4$ & $0.0081$ & $2.906$ & $0.0046$ & $2.718$ & $0.0033$ & $2.567$ \\
$2f_0 -f_\mathrm{m}'$ & $5.027272$ & $29.86$ & $-12.99$ & $-99.47$ & $4.4$ & $-1.5$ & $-5.9$ & $0.0051$ & $0.709$ & $0.0039$ & $0.484$ & $0.0020$ & $0.600$ \\
$3f_0 -f_\mathrm{m}'$ & $7.541893$ & $7.10$ & $-4.20$ & $-48.85$ & $0.8$ & $-0.5$ & $-3.2$ & $0.0038$ & $5.460$ & $0.0039$ & $5.365$ & $0.0023$ & $5.214$ \\
$f_\mathrm{m}'$ & $0.001970$ & $110.33$ & $196.07$ & $195.70$ & $11.4$ & $12.4$ & $9.3$ & $0.0036$ & $0.717$ & $0.0022$ & $0.539$ & $0.0016$ & $0.127$ \\
\hline
\end{tabular}
\end{minipage}
\end{table*}

\begin{figure}

 \includegraphics[height=8.1cm,angle=-90]{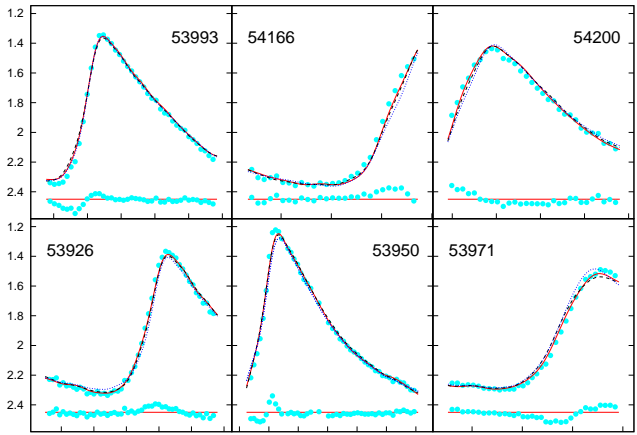}
  \caption{$V$ light curves and the triplet (dotted blue line), quintuplet (dashed black line), and light curve 
solution D (solid red line) fits to the data are shown on some representative nights. On the X axes tick marks 
are at 0.005 d on each plot. Bellow the light curves the residuals if light curve solution D is removed are shown. 
The residuals show smooth light curves with systematic deviations from zero. Large residuals appear 
at around minimum and maximum light and on the rising branch. However, as e.g., on JD 2454200 the residual is 
continuously small but negative on the upper part of the descending branch. These systematic deviations explain 
the large rms. scatter of the residual light curve.}
  \label{lcfit}
\end{figure}

\begin{figure}
  \centering
  \includegraphics[width=8.8cm]{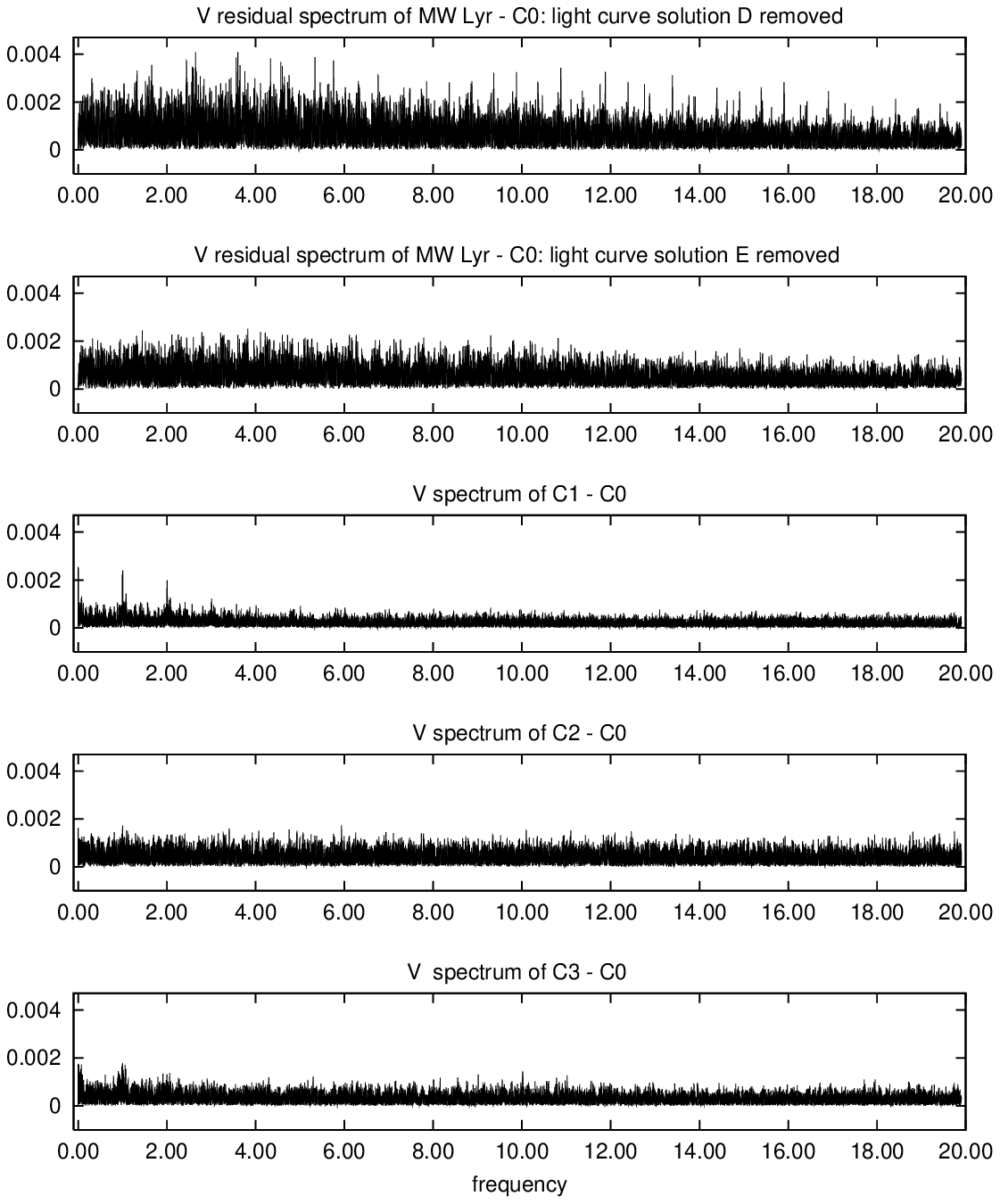}
  \caption{Comparison of the residual spectra of the $V$ light curve of MW~Lyr with the spectra of the check 
stars' light curves. The identifications and magnitudes of the C0 comparison and C1, C2, C3 check stars are 
given in Table~\ref{comps}. Two residual spectra of MW~Lyrae are shown. After removing the 66 frequencies of 
light curve solution D significant peaks in the spectrum still appear (see also Fig.~\ref{residualszin}). 
Removing further 30 frequencies from the data the residual spectrum is still at 2-3 times higher level than the 
noise spectrum of the check stars. As the brightness of MW~Lyr varies between 13.10 mag and 14.30 mag, while 
C1, C2, and C3 have brightnesses of 12.99, 13.92 and 14.36 mag, respectively, brightness differences cannot 
account for the large residual signal of MW~Lyrae.}
  \label{residualoh}
\end{figure}

In Cols 9-14 of Table~\ref{big} the amplitudes and phases of the $B$, $V$, and $I_\mathrm{C}$ light curves 
according to sine term decomposition are given. Frequencies with S/N  (S/N is defined as the ratio of the 
$V$ amplitude and the mean value of the residual spectrum of light curve solution D in the vicinity of the 
given frequency) smaller than 3 in each band are denoted by asterisks. The inclusion of these frequencies 
in the light curve solution has minimal effect on the results. However, as we are focusing on how accurately 
the light curve of a Blazhko variable can be fitted with the mathematical model of equidistant modulation 
side frequencies, we decided to include these low S/N signals also in the frequency solution.

The rms scatter of the residual light curves of light curve solution D in the $B$, $V$, and $I_\mathrm{C}$
bands are 0.026, 0.020 and 0.015 mag, respectively. Comparing these rms values to the residuals of the 
comparison -- check stars' light curves, which are in the 0.008 -- 0.015 mag range it seems that the residual 
scatter of MW~Lyrae is significantly larger than expected. The $V$ magnitudes of MW~Lyr vary between 12.95 
and 14.35 mag, the brightnesses of the check stars C1, C2, and C3 nearly equal to the variables maximum, 
mean and minimum  brightnesses, respectively (see the magnitudes and colours of the comparison and check 
stars in Table~\ref{comps}). The mean $\Delta(B-V)$  between MW~Lyr and C0 is $-0.25$ mag, while the 
$\Delta(B-V)$ colour differences of the three check stars are between 0.02 and 0.28 mag. Neither the brightness 
nor the colour differences can account for the larger scatter of the residual light curve of MW~Lyr.

Fig.~\ref{lcfit} shows some examples how light curve solutions A, B, and D fit the observations on different 
nights. It can be seen that the observations deviate systematically even from the fit of solution D that 
involves 66 frequency components. The deviations are systematic, they are centred on the minimum-rising 
branch-maximum phase of the pulsation with different shapes. The residual mimics the behaviour of the large 
amplitude modulation on a much smaller scale, but without any definite  periodicity.

Just for a trial we have included further 30 frequencies in the fit (consecutive prewhitening with the highest 
peaks appearing in the spectra shown in Fig.~\ref{residualszin}, light curve solution E) to see whether these 
systematic deviations can or cannot be eliminated with further distinct frequency components. Though the rms of 
e.g., the $V$ light curve has been reduced to 0.018 mag this way, the systematic deviations of the light curves 
shown in Fig.~\ref{lcfit} have hardly decreased. 

Fig.~\ref{residualoh} compares the residual spectra of the $V$ light curve prewhitened with light curve 
solution D and E with the spectra of the  $V$ light curves of the  C1, C2 and C3 check stars. The differences 
are striking. The mean level of the residual spectrum of light curve solution E is still $2-3$ times larger than 
that of the spectra of the check stars.

Most probably stochastic and/or chaotic behaviour of the modulation  bring forth the enhanced residual signals.
It seems that the light curves of large modulation amplitude Blazhko variables cannot be modelled with the 
required accuracy with the Fourier sum of finite number of frequency components.

\subsection{The maximum brightness -- maximum phase variation}

\begin{figure}
  \centering
  \includegraphics[width=8.8cm]{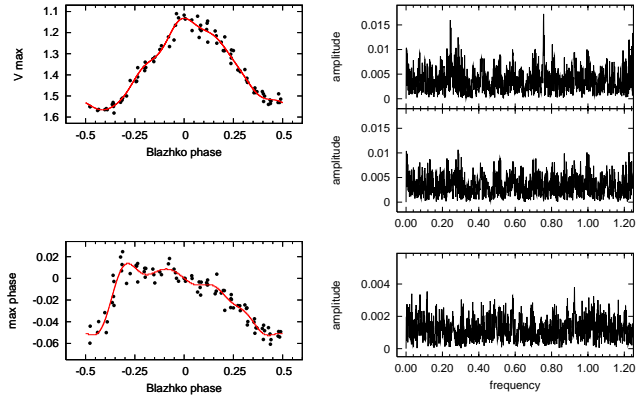}
\caption{Maximum $V$ brightness and maximum phase values vs. Blazhko phase are plotted in the top-left and 
bottom-left panels, respectively. Sixth order Fourier fits to the data are also drawn in these plots. The data 
are prewhitened with these fits. The right-top panel shows the amplitude spectrum of the residual of the maximum 
brightness data. The highest peak appears at $f=0.7559$ cd$^{-1}$, which equals to $12.5 f_\mathrm{m}$ within the 
uncertainty. Prewhitening the data also with this frequency, there is no further significant peak in the residual 
spectrum (middle-right panel). The residual spectrum of the prewhitened maximum phase data (bottom-right panel) 
does not show the $f=0.7557$ cd$^{-1}$ signal. There is no peak appearing at the same frequency in the maximum phase 
residual (prewhitened with the modulation frequency and its harmonics) and the maximum brightness residual 
(prewhitened with the modulation frequency and its harmonics plus $12.5\,f_\mathrm{m}$) spectra. }
  \label{max-oc}
\end{figure}

\begin{figure*}
  \centering
  \includegraphics[width=17.5cm]{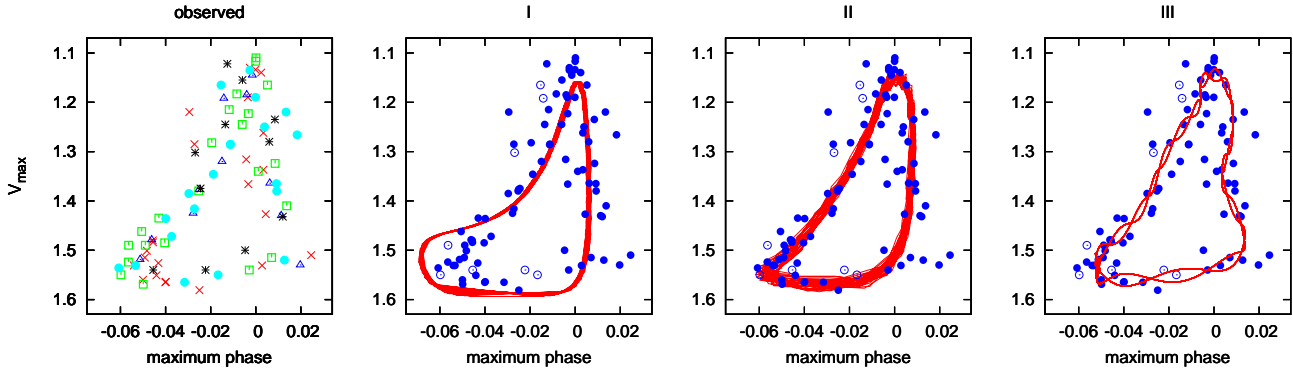}
\caption{$V$ maximum brightness  vs. maximum phase data of the 88 observed maxima. Data for the consecutive 50-70 
days segments (3-5 Blazhko cycles) are shown by different symbols in the left panel. No systematic change with time 
in the $V_\mathrm{max}-Phase_\mathrm{max}$ plot is evident. For comparison, in the I, and II panels synthetic $V_\mathrm{max}-Phase_\mathrm{max}$ curves according to the light curve solution A (triplet) and D (that involves 
the septuplet solution of the modulation plus 4 modulation components with $12.5\,f_\mathrm{m}$ separation, and 
4 modulation components with $f_\mathrm{m}'=0.0019$\,cd$^{-1}$ modulation frequency) are plotted. In panel III the 
fitted curve corresponds to the harmonic fits to the maximum brightness and maximum phase data with the frequencies 
given in Table~\ref{max-octab}. The observed $V_\mathrm{max}-Phase_\mathrm{max}$  data are overplotted on the 
synthetic solutions, the open symbols denote less accurate data. The scatter of the observations is much larger 
than observational uncertainties can explain as data are accurate within $\pm0.02$ mag and $\pm 0.003$\,d 
$\pm0.0075$ phase ranges.}
  \label{egg}
\end{figure*}

The modulation of the pulsation light curve of MW~Lyr can be also followed using the maximum brightness, 
maximum phase data given in Table~\ref{maxtb}. The maximum brightness and maximum phase values of 
the $V$ light curve are shown in the left panels of Fig.~\ref{max-oc} folded with the modulation period. 
None of these plots can be fitted  with a single sine wave. The maximum phase data show a highly asymmetric 
shape, while there is a bump on the rising branch of the maximum brightness data. For an accurate fit, both 
plots need at least 6th order Fourier sums. Table~\ref{max-octab} lists the amplitudes of the 6th order Fourier 
components of the maximum brightness and maximum phase fits. The amplitudes change with increasing order, 
they follow a similar trend for the maximum brightness data as the amplitudes of the sidelobe frequencies of 
the Fourier spectrum of the light curve do: the 1st, 2nd and 4th order components have pronounced amplitudes while 
the amplitudes of the 3rd, 5th and 6th order components are small. On the contrary, the amplitudes of the 3rd, 
4th and 5th order components of the maximum phase fit are very similar. The 0.022 mag and 0.0066 phase 
(0.0026~days) rms scatter of the residuals of the maximum brightness and maximum phase fits are, however, too 
large as we estimate that the data are accurate within $\pm0.02$ mag and $\pm0.0025$ d, respectively.

The residual spectra of the maximum brightness and maximum phase data prewhitened with the 6th order fits of 
the modulation frequency are shown in the right panels of Fig.~\ref{max-oc}. The residual spectrum of the 
maximum brightness data shows a peak with 0.015 mag amplitude at 0.7559 cd$^{-1}$. This frequency equals within 
the uncertainty to the 0.7555 cd$^{-1}$ frequency value of $12.5f_\mathrm{m}$. Though this modulation component 
itself does not appear in the spectrum of the light curve, it has measurable amplitudes at the 
$kf_0\pm12.5f_\mathrm{m}$ sidelobe positions as discussed in the previous Section. There is no sign of this 
modulation component in the maximum phase data. Accordingly, the modulation connected to the $12.5f_\mathrm{m}$
frequency is dominantly amplitude modulation.

The removal of the $12.5f_\mathrm{m}$ component from the maximum brightness data lowers the residual scatter 
to 0.020 mag only, which is still too high to be explained with observational inaccuracy. Both the maximum brightness 
and maximum phase data reflect the imperfection of modelling the modulation using the Fourier sum of discrete 
frequency components. Most probably the observations cannot be traced with the expected accuracy using finite 
number of strictly periodic signals.

Plotting the maximum brightness vs. maximum phase data, as Fig.~\ref{egg} shows, significant scatter appears 
around a triangular shape curve. No evolution of the data with time account for the scatter, 
data from different segments of the observations are equally scattered without any systematics. Synthetic 
maximum brightness and maximum phase curves are drawn according to the triplet light curve solution, light curve 
solution D, and maximum brightness and maximum phase fits involving 6 harmonics of the modulation and  
$12.5 f_\mathrm{m}$  for the maximum brightness data in the panels I, II and III  in Fig.~\ref{egg}, respectively.
Many observed maximum data are out of the ranges of any of the model fits, that cannot be explained by data inaccuracy.

\begin{table}
\caption{Fourier amplitudes of the fits to the maximum brightness and maximum phase data}
 \label{max-octab}
  \begin{tabular}{rcccc}
  \hline
frequency&\multicolumn{1}{c}{maximum brightness}&\multicolumn{1}{c}{maximum phase}\\
& Amp [mag] &Amp [pulsation phase] \\
 \hline
 $f_\mathrm{m}$ &0.2044& 0.0293\\
 $2f_\mathrm{m}$&0.0158& 0.0113\\
 $3f_\mathrm{m}$&0.0063& 0.0041\\
 $4f_\mathrm{m}$&0.0112& 0.0047\\
 $5f_\mathrm{m}$&0.0058& 0.0034\\
 $6f_\mathrm{m}$&0.0042& 0.0016\\
 $12.5f_\mathrm{m}$&0.0125& $-$\\
\hline
&\multicolumn{1}{c}{rms$ = 0.0203$}&\multicolumn{1}{c}{rms$ = 0.0066$}\\
\hline
\end{tabular}
\end{table}

\subsection{The mean light curve}

\begin{figure}
  \includegraphics[height=4.2cm]{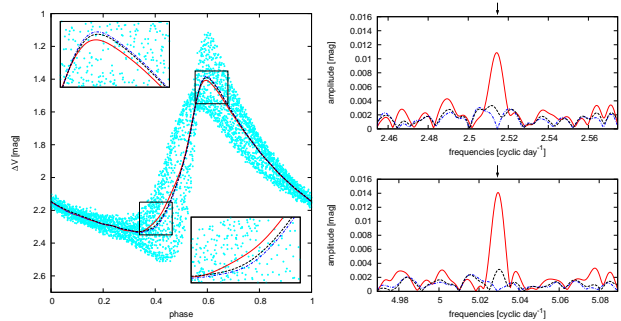}
  \caption{Comparison of the mean light curves defined as $a$) the Fourier fit of the pulsation frequency and its 
harmonics to the $V$ observations from the first season (solid red line), $b$) the Fourier parameters of the 
pulsation components are taken from a triplet solution of the data of the first season (dashed black line), $c$) the 
Fourier parameters of the pulsation components are taken from light curve solution D of the entire dataset 
(blue dot-dashed line). The residual spectra in the vicinity of the pulsation frequency and its first harmonic  
component are shown in the right panels. In method $a$ the modulation side components are also removed but in a 
second step after the pulsation frequencies have been already removed from the data. Note the smaller amplitude of 
the fitted light curve and the high residual signals at $f_0$ and $2f_0$ when applying method $a$. }
  \label{f5}
\end{figure}

Interesting questions arise in connection with the mean light curves of variables showing light curve modulation.
Namely, we still do not know whether \\
a) the mean light curve is or is not the same as the light curve of a star with the same physical parameters but 
not showing Blazhko modulation,\\
b) the actual shape of the light curve in any phase of the modulation corresponds to  the mean light curve.

Our extended and accurate data make it possible to find the correct answers to these questions.
However, first we have to define the mean pulsation light curve of Blazhko variables correctly.

The simplest way to define the mean light curve is to fit all the data with a high enough order Fourier sum of 
the pulsation frequency and its harmonics (method $a$). However, this procedure may yield a mean light curve that 
is biased by the uneven data sampling, even in the case of an extended data set. The typical procedure of light curve 
analysis of Blazhko stars is  prewhitening first with the pulsation frequency and its harmonics (with the mean 
light curve defined above), and then, the prewhitened data are analysed to identify the modulation components. 
As a result of the incorrect shape of the removed mean light curve during this procedure false signals in the 
vicinity of the pulsation components emerge. A better approximation of the mean light curve is gained by a 
fit that takes the pulsation and modulation frequency components simultaneously into account. The mean pulsation 
light curve according to a triplet solution light curve fit (method $b$) may significantly differ from the 
mean light curve defined by method $a$.

Fig.~\ref{f5} demonstrates the differences between the fits and the residual spectra if the data are fitted and 
prewhitened by the pulsation and modulation components in consequtive steps and simultaneously. In this Figure 
data from the first season of the observations are shown. This data set, which contains 3900 data points from 
120 nights is, however,  more dense and extended than any previous photometric observation of a Blazhko variable.
Though these data cover the whole pulsation light curve in each 0.05 phase of the modulation, the differences 
between the mean light curves defined by method $a$, and by method $b$ are significant.

The fitted mean light curves shown in Fig \ref{f5} are:
\begin{enumerate}
\item{ the Fourier fit to the data with the pulsation frequency and its harmonics ({\it method a}); }
\item{the fit taken from the simultaneous triplet frequency solution of the data ({\it method b});  }
\item{the pulsation light curve taken from the complete light curve solution of the entire data set (light curve 
solution D as given in Table~\ref{big}).}
\end{enumerate}

Method $a$ gives substantially smaller amplitude around minimum and maximum light due to the oversampling of 
the small amplitude phase of the modulation in the data, than the fits  either from method $b$ or from the 
light curve solution D. The latter fits are very similar, showing that with the inclusion of the most prominent 
modulation components (triplets) in the light curve solution quite a reliable mean light curve can be gained 
even if the data sampling is biased.

The right panels in Fig.~\ref{f5} show the residual spectra of the three fits in the vicinity of the pulsation 
frequency and its first harmonic. In order to make the residuals comparable, the modulation side frequencies are 
also removed in a second step in method $a$. These residual spectra show high amplitude signals at $f_0$ and $2f_0$,
which also points to the inadequacy of prewhitening the data in consequtive steps. 

We thus conclude that one has to be cautious how to define the mean light curve of a Blazhko variable as 
from method $a$ and method $b$ different results emerge. In real observations data sampling is always somewhat 
unevenly distributed. As a consequence, the mean light curve defined without taking the modulation components 
also into account, may give an incorrect result for variables showing any type of light curve modulation.

From here on, we use the Fourier parameters of the pulsation components given in Table~\ref{big} to define 
the mean pulsation light curve.

\subsection{Light curve changes during the Blazhko cycle}

\begin{table*}
 \caption{Fourier parameters of the $V$ light curves in the 0.05 phase bins of the modulation}
\label{20vfour}
  \begin{tabular}{c@{}rcrclllllrrrrr}
  \hline
Bl phase&Order&rms&$<V>$&N&$\Phi(f_0)$&$\Phi_{21}$&$\Phi_{31}$&$\Phi_{41}$&$\Phi_{51}$&$A(f_0)$&$A(2f_0)$&$A(3f_0)$&$A(4f_0)$&$A(5f_0)$\\
&&&&&[rad]&&&&&\multicolumn{5}{c}{[mag]}\\
 \hline
$0.00-0.05$&14&0.015&  2.031 &228&1.691   &2.423   &5.173   &1.483   &4.146   &0.459   &0.244   &0.148   &0.092   &0.059\\
$0.05-0.10$&14&0.024&  2.024 &277&1.737   &2.430   &5.136   &1.465   &4.166   &0.460   &0.247   &0.142   &0.087   &0.052\\
$0.10-0.15$&13&0.032&  2.017 &294&1.792   &2.417   &5.095   &1.364   &3.980   &0.457   &0.234   &0.130   &0.077   &0.047\\
$0.15-0.20$&13&0.020&  2.011 &347&1.866   &2.438   &5.024   &1.342   &3.824   &0.441   &0.209   &0.108   &0.060   &0.033\\
$0.20-0.25$&13&0.019&  2.006 &285&1.965   &2.449   &5.030   &1.285   &3.800   &0.421   &0.197   &0.087   &0.042   &0.019\\
$0.25-0.30$&13&0.026&  2.005 &341&2.053   &2.452   &5.018   &1.462   &4.074   &0.394   &0.176   &0.078   &0.041   &0.018\\
$0.30-0.35$&13&0.026&  2.003 &352&2.155   &2.416   &5.023   &1.426   &4.278   &0.369   &0.156   &0.066   &0.035   &0.014\\
$0.35-0.40$&12&0.022&  2.006 &376&2.210   &2.386   &5.067   &1.597   &4.371   &0.349   &0.149   &0.059   &0.031   &0.014\\
$0.40-0.45$&11&0.019&  2.007 &280&2.274   &2.440   &4.978   &1.600   &4.114   &0.333   &0.132   &0.051   &0.023   &0.012\\
$0.45-0.50$& 9&0.026&  2.003 &147&2.266   &2.501   &4.913   &1.789   &4.262   &0.331   &0.127   &0.059   &0.023   &0.008\\
$0.50-0.55$& 9&0.016&  1.999 &194&2.309   &2.453   &4.858   &1.543   &4.501   &0.308   &0.118   &0.049   &0.019   &0.008\\
$0.55-0.60$& 9&0.016&  2.000 &303&2.240   &2.446   &4.906   &1.196   &3.270   &0.296   &0.109   &0.049   &0.010   &0.002\\
$0.60-0.65$& 9&0.022&  1.993 &381&2.083   &2.445   &5.046   &1.115   &3.910   &0.312   &0.103   &0.052   &0.016   &0.005\\
$0.65-0.70$& 6&0.018&  1.990 &338&1.938   &2.433   &5.104   &1.136   &3.755   &0.333   &0.117   &0.052   &0.018   &0.007\\
$0.70-0.75$& 8&0.027&  1.995 &192&1.816   &2.443   &5.054   &1.073   &3.838   &0.355   &0.146   &0.065   &0.035   &0.016\\
$0.75-0.80$& 9&0.027&  2.004 &288&1.770   &2.323   &5.016   &1.283   &3.729   &0.368   &0.177   &0.093   &0.054   &0.033\\
$0.80-0.85$&10&0.025&  2.009 &376&1.737   &2.346   &5.075   &1.342   &3.850   &0.392   &0.189   &0.102   &0.068   &0.037\\
$0.85-0.90$&11&0.018&  2.012 &268&1.681   &2.392   &5.105   &1.467   &4.096   &0.416   &0.215   &0.121   &0.077   &0.043\\
$0.90-0.95$&13&0.025&  2.028 &277&1.650   &2.386   &5.196   &1.501   &4.220   &0.442   &0.239   &0.140   &0.080   &0.055\\
$0.95-1.00$&15&0.015&  2.026 &252&1.676   &2.447   &5.141   &1.453   &4.094   &0.462   &0.250   &0.154   &0.094   &0.065\\
\hline
\multicolumn{15}{c}{average values}\\
\hline
$0.00-1.00$ & 11&0.022&  2.008      &  290 &1.946   &2.423  & 5.046&   1.395&   4.014&   0.385&   0.177&   0.090&   0.049&   0.027\\
\hline
\multicolumn{15}{c}{parameters of the light curve solution given in Table~\ref{big} (mean light curve)}\\
\hline
$0.00-1.00$ & 12&0.020&  1.966     & 5796 &1.919   &2.262  & 4.878&   0.943&   3.239&   0.374&   0.159&   0.076&   0.037&   0.020\\
\hline
\multicolumn{15}{c}{parameters of the mean light curve of the time transformed data}\\
\hline
$0.00-1.00$ & 13&0.020&  2.008     & 5796 &1.955   &2.417  & 5.068&   1.405&   4.034&   0.384&   0.176&   0.089&   0.049&   0.027\\
\hline
\end{tabular}
\end{table*}

\begin{figure*}
  \includegraphics[width=8.5cm,angle=-90]{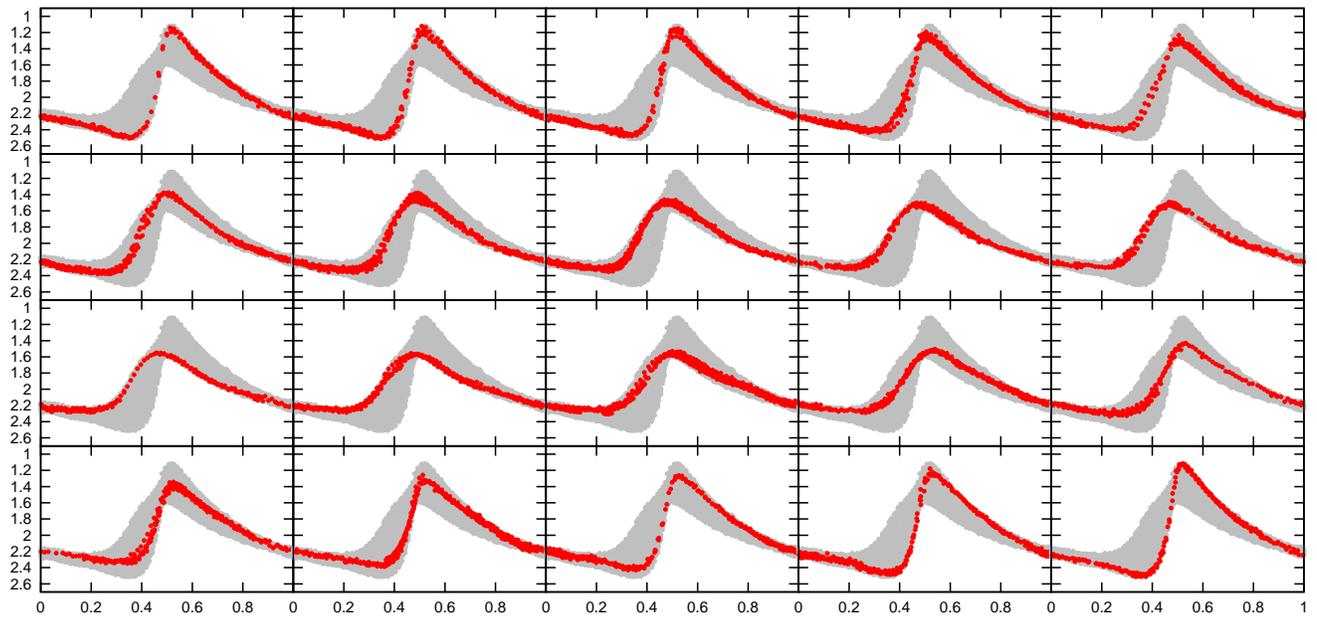}
  \caption{$V$ light curves of MW~Lyrae in 20 different phases of the Blazhko cycle are shown. The folded light 
curve of the complete data set is shown in gray colour.}
  \label{lcossz}
\end{figure*}

\begin{figure}
  \includegraphics[height=7.8cm]{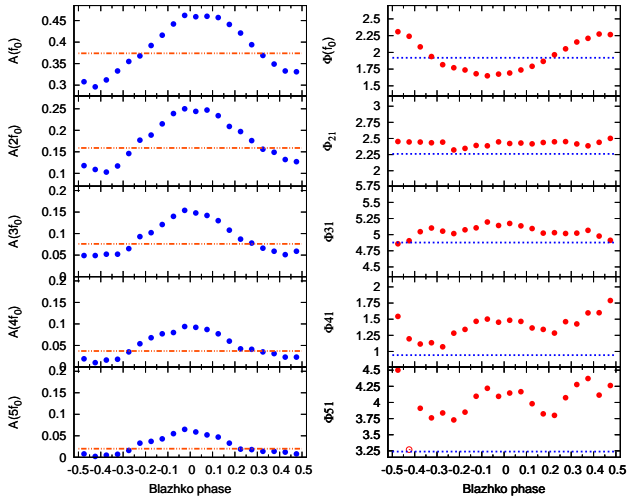}
  \caption{Fourier parameters of the $V$ light curves in 20 phase bins of the Blazhko cycle are plotted. The open 
circle in the  $\Phi_{51}$ plot denotes a very uncertain data point, the amplitude of the $5f_0$ component is only 
0.002 mag in this phase bin. The amplitudes and the phases  are plotted on the same scales. Note that the Fourier 
phases have opposite sign as the directly measured phase of the maximum brightness  shown in Fig~\ref{max-oc}.
The amplitudes of the $kf_0$ ($k=1,...5$) components and the $\Phi(f_0)$ phase show smooth, sinusoidal variations
with the smallest phase value when the amplitudes are the highest and with the largest phase value when the 
amplitudes are the smallest. On the contrary the epoch independent phase differences ($\Phi_{k1}$) show 
surprisingly different  behaviour. There are small if any changes  in $\Phi_{21}$, and the variation in  
$\Phi_{31}$ is also very small taking into account the significant changes in the light curves' shapes as 
shown in Fig.~\ref{lcossz}. $\Phi_{41}$  and $\Phi_{51}$ vary during the Blazhko cycle showing complex changes   
on the time scale of about half the modulation period. For comparison, Fourier parameters of the mean light 
curve as given in Table~\ref{big} are also drawn by dashed lines in the plots.
While the $\Phi(f_0)$ value of the mean light is  close to the average of their observed values in the different 
Blazhko phases, the phase differences in each phase of the modulation are larger than the phase differences of the 
mean light curve. The Fourier amplitudes of the mean light curve are $0.01-0.02$ mag fainter than the averages of 
their values in different phases of the Blazhko cycle.}
  \label{param}
\end{figure}

\begin{figure}
  \includegraphics[width=9cm]{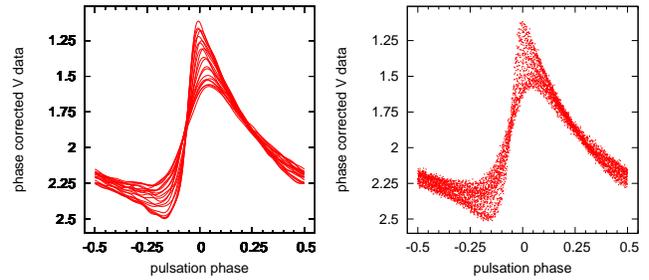}
  \caption{
{\it Left panel:} Fitted curves of the $V$ light curve of MW~Lyr in 20 bins of the Blazhko cycle. Each light curve is 
phase shifted with  the $\Phi(f_0)$ phase value. {\it Right panel:} Time transformed $V$ light curve of MW~Lyrae. 
The time transformation was defined from a second order harmonic fit to the  $\Phi(f_0)$ phases in different phases 
of the modulation.  }
  \label{megold}
\end{figure}

\begin{figure}
  \includegraphics[width=7cm]{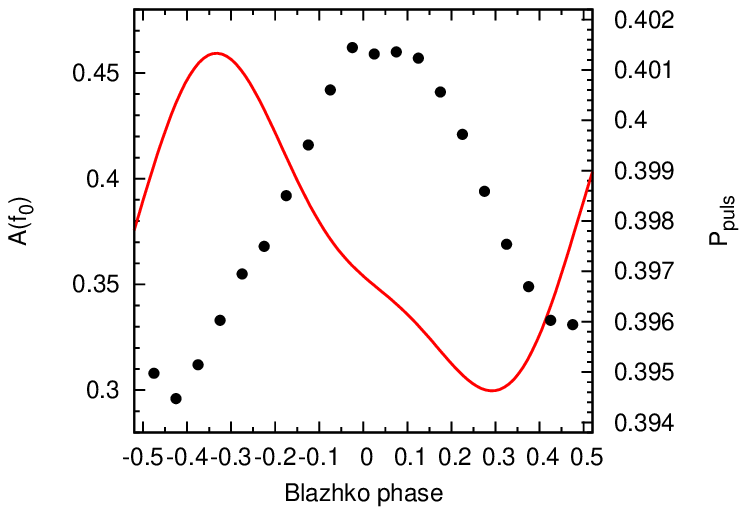}
  \caption
{Phase relation of the amplitude (dots) and period (red line) changes of MW Lyrae during the Blazhko cycle. The amplitude  
and period variations are measured as the amplitude and as the derivative of the phase variation of the $f_0$ pulsation 
frequency component, respectively.}
  \label{a-p}
\end{figure}

Utilizing the full coverage of the pulsation period in each phase of the modulation in our data set we can reliably compare 
the mean light curve to the light curves in different phases of the Blazhko modulation. In Fig.~\ref{lcossz} 
the $V$ light curves of MW~Lyr are shown for 20 bins of the modulation cycle. The scatter of these light curves can be 
partially explained by the regular light curve changes that take place even in 0.05 phase intervals of the modulation. 
Modulation frequencies that are not integer multiplets of $f_\mathrm{m}$ (e.g., $12.5f_\mathrm{m}$ and $f_\mathrm{m}'$) 
also result in enhanced scatter of the data phased with the modulation period. Moreover, our experience, that the light 
curve cannot be fitted with the required accuracy supposing regular modulations with different modulation periods as 
discussed in Sect 3.1 and 3.2, means that this irregular character of the modulation adds some extra noise to the light 
curves in the different phase bins, as well.

Notwithstanding these effects, the light curves in the different phase bins are well defined, and can be characterized 
by the Fourier amplitudes and phases of the pulsation frequency and its harmonics.

The changes in the Fourier amplitudes and phases (phase differences) of $f_0$ and its lower harmonics and also the 
average values of these parameters are listed in Table~\ref{20vfour}. For comparison, the last line gives the 
corresponding parameters of the mean $V$ light curve. The first five columns in  Table~\ref{20vfour} give the phase bin, 
the order of the Fourier sum fitted to the data, the residual scatter of the fit, the intensity weighted mean magnitude, 
and the number of data points belonging to the given bin, respectively. The phases of the $f_0$ pulsation frequency can 
be read from the sixth column (initial epoch is 2\,453\,887.00) while the next four columns list the epoch independent 
phase differences $\Phi_{k1}$ ($k=2,...5$). The amplitudes of the $kf_0$ ($k=1,...5$) components are given in the last 
five columns. The residual scatters of the fitted harmonic functions to the data in the different phase bins of the 
modulation are within the 0.015-0.032 mag range. The average  of the rms values of the fits is 0.022 mag, slightly 
larger than the rms of the light curve solution D of the entire data set.

The data listed in Table~\ref{20vfour} and plotted in Fig.~\ref{param} show that there are major differences between the 
Fourier parameters of the mean light curve (the Fourier parameters of the $kf_0$ components of the full light curve 
solution given in Table~\ref{big}) and the average values of the same Fourier parameters of the light curve fits in 
different phases of the Blazhko cycle. The amplitudes of the  $kf_0$ components of the mean light curve are 
systematically $0.01-0.02$~mag smaller than the averages derived from the fits of the individual light curves. 
The $\Phi_{21}$, $\Phi_{31}$, $\Phi_{41}$ and $\Phi_{51}$ phase differences of the mean light curve solution are 
0.16, 0.17, 0.35 and 0.78 rad smaller than the averages of the corresponding parameters of the light curves in 
different phases of the Blazhko cycle.

These large differences make it unambiguous that in each phase of the Blazhko modulation the light curve of MW Lyr 
differs from its mean light curve.

The $\Phi_{k1}$ phase differences show quite a surprising behaviour. In spite of the large amplitude of the Blazhko 
modulation of MW Lyrae, $\Phi_{21}$, and $\Phi_{31}$ hardly vary, while the changes in $\Phi_{41}$ and $\Phi_{51}$  
show complex behaviour with cycle length half of the modulation cycle. For the higher order components the amplitudes at 
around Blazhko minimum are so small that the errors in the phases become too large to make any firm conclusion about 
their variations. 

The light curve changes during the Blazhko cycle can be summarised as follows:
\vskip 3pt

\noindent 
{\it Light curve changes connected to the amplitude variations:}

The amplitudes of the  $kf_0$ frequencies show parallel changes but the amplitudes of the higher order components 
decrease more drastically than the amplitudes of the lower order components. For example, the amplitude ratio 
$A(f_0)/A(f_5)$ is about 7 in the three largest amplitude phase bins, but  this ratio is  as large as 50-150 for 
the three smallest amplitude phase bins. 

The light curves at around maximum amplitude phase of the modulation can be fitted accurately with 10-15 harmonic 
components of $f_0$. The light curves are much more sinusoidal in the small amplitude phases, they can be fitted 
with 6-9 harmonic components with the required accuracy. 
\vskip 3pt

\noindent
{\it Light curve changes connected to the phase variations:}

The phases of the lower order harmonic components show very harmonized changes as indicated by the small variations 
of the lower order phase differences. This means that the changes in the light curves' phase  can be characterized  
basically by one parameter, with the phase of the $f_0$ pulsation frequency, $\Phi(f_0)$.

The $\Phi_{41}$ and $\Phi_{51}$ phase differences show double wave curves indicating that the time scale of their 
variation is about half of the period of the  modulation. The appearance of the $2f_\mathrm{m}$, $4f_\mathrm{m}$  
frequency components in the Fourier spectrum of the whole data set is probably connected with this double periodic 
behaviour of the higher order phase differences. 

We have also tested how accurately the observations can be fitted with a simple mathematical model which 
describes the amplitude and phase changes of the different order pulsation components with different order Fourier series.
We have found that such a light curve solution fails to fit the observaions with 
similar accuracy as the Fourier sum of the pulsation and modulation side lobe frequencies involving even  
a smaller number of parameters. This result means that the modulation has a very complex behaviour that
could not be described by a mathematical model of amplitude and phase modulations of harmonic functions.

Phenomenologically, amplitude modulation occurs if there are observed changes in the brightnesses of the  
light curve maxima, while phase modulation manifests itself as a missing fix point on the rising branch of the 
folded light curve.

Exploiting the slight changes in the lower order phase differences, we can `harmonize' the phases of the light 
curves with a simple phase correction. Shifting the fitted light curves of the different bins by their $\Phi(f_0)$ 
values a surprisingly coherent light curve series emerge as shown in Fig.~\ref{megold}. The pronounced fix point 
occurring on the rising branch  in this figure validates our simple treatment separating the phase modulation of 
the light curve from the amplitude modulation by correcting the phases with the $\Phi(f_0)$ values. 

A similar procedure can be applied on the whole data set, as well. Fitting the $\Phi(f_0)$ values of the 20 phase 
bins with a 2nd order Fourier sum we can define a continuous function of the phase variation. 
It is supposed that by transforming the times of the observations according to this function we `get rid of' 
the phase modulation component of the modulation.
As the right panel in Fig.~\ref{megold} shows, this is indeed the case, the light curve of 
the time transformed data shows very regular amplitude modulation with a pronounced fix point on its rising branch.
The phase difference of the amplitude peaks of the time transformed data
is much smaller than in the original data.
Its range is consistent with the value expected from the amplitude 
variation.

Without starting the light curve analysis from the beginning using this phase corrected data set, we have 
only checked 
\begin{itemize}
\item whether or not the same frequency components occur in the spectrum of the time transformed data as in 
the original data set,

and 
\item how the rms scatter of the time transformed data compares to the rms residual of the original data.
\end{itemize}

All the frequencies but $2f_0-4f_m$ and $3f_0-4f_m$ listed in Table~\ref{big} can be identified in the residual 
spectrum of the time transformed data. The $f_m$ and $2f_m$ modulation sidelobe components can be detected up to 
higher order harmonics as contrasted with the spectrum of the original data. 

The residual scatter of the original $V$ light curve was 0.020 mag, the rms scatter of the time transformed data 
remains the same, 0.020 mag if the modulation components up to the appropriate order are taken into account. 
The Fourier parameters of the mean pulsation light curve of the time
transformed data (given in the last line in Table~\ref{20vfour}) equal within their error
ranges with the average values of the Fourier  parameters of the observed
light curves in different phases of the Blazhko cycle.
The residual light curve of the time transformed data  also shows large deviations at around minimum-rising branch-maximum 
phases of the pulsation. 

Without finding the correct explanation of the Blazhko phenomenon we cannot decide which characterization of the 
light curve is correct: the Fourier analysis of the light curve as it is, or the separation of the phase and amplitude 
modulation components of the light curve modulation by an appropriate transformation of the phases (times) of the 
observations. The simplicity and the low number of independent parameters involved in the time transformation applied 
suggest that this new treatment of the light curve may lead to a headway in the study of Blazhko variables.

We also remark that the time transformation can be explained as a continuous change in the pulsation period during 
the Blazhko cycle as \cite{stothers} has also interpreted. The full range of period change determined from the 
derivative of the phase shift curve is 0.006~days, i.e. $\delta P/P=0.015$. The period of the pulsation is about 
0.401~days around Blazhko phase 0.65 and 0.395~days around Blazhko phase 0.30 as Fig.~\ref{a-p} shows.  
The amplitude of the period variation is somewhat larger than \cite{stothers} derived for RR Lyrae itself, 
but keeping in mind that for RR Lyr temporal periods were determined for some days long intervals of the 
observations, most probably a reduced value of its real period change was found.

\section{Conclusions}

The photometric observations of MW~Lyr analyzed in this paper comprise the most extended and accurate data set of a 
Blazhko variable ever obtained. Utilizing this unique opportunity a detailed and circumspect phenomenological description 
of the modulation is given, which may provide  crucial information  to find the correct explanation of the phenomenon.

The main results and conclusions of the analysis are the followings:

\begin{itemize}

\item{In the Fourier spectrum of the light curve besides the $kf_0\pm f_\mathrm{m}$ triplet frequencies 
$kf_0\pm 2f_\mathrm{m}$ quintuplet, and  $kf_0\pm 4f_\mathrm{m}$ septuplet components also appear.}

\item{Both $f_\mathrm{m}$ and $2f_\mathrm{m}$ frequencies can be detected in the spectrum.}

\item{Frequency components at $kf_0\pm 12.5f_\mathrm{m}$ are  detected. If these frequencies are not `just by chance' 
at $\pm 12.5f_\mathrm{m}$ separations but somehow they are indeed connected to the main modulation frequency, then 
it is a great challenge to find an answer to their origin. }

\item{Modulation with $12.5f_\mathrm{m}$ frequency can also be detected in the maximum brightness data but not in 
the maximum phase observations. Consequently, the modulation connected to the $12.5f_\mathrm{m}$ frequency is dominantly 
amplitude modulation. }

\item{\cite{stothers} mentioned the lack of large amplitude modulation around minimum phase of the pulsation as a 
failure of his model. The amplitude of the modulation around minimum phase of the pulsation in MW Lyr is, however, 
commensurable with the amplitude of the modulation in maximum brightness. On the contrary, the modulation of MW Lyrae is 
more strictly periodic and regular than it would be expected to be if the triggering mechanism behind the modulation were 
the cyclic weakening and strengthening of the turbulent convection in the ionization zones as \cite{stothers} proposes. }

\item{Though the modulation shows high degree of regularity both in the phase (period) and in the amplitude changes, 
the light curve cannot be fitted with the required accuracy even with 66 identified and 
further 30 frequencies appearing in the residual spectrum. Significant deviations in the residual light curve are 
concentrated at the minimum-rising branch-maximum phase of the pulsation. These residuals, however, do not show any 
periodicity, most probably they can be explained with some stochastic and/or chaotic behaviour of the modulation itself.}

\item{The mean pulsation light curve defined by the Fourier parameters of the $kf_0$ pulsation frequency components of 
the full light curve solution differs significantly from the light curve in any phase of the modulation. Especially 
the $\Phi_{k1}$ phase differences of the mean light curve are discrepant, they are out of the range of the phase 
difference values measured in any phase of the Blazhko cycle.}

\item{The light curves in the small and large amplitude phases of the modulation can be fitted with 6-10 and 11-15 
order harmonic fits, respectively. If nonlinear effects (e.g., shock waves) account for the occurrence of the higher 
order harmonic components of the pulsation, then their diminishing amplitudes in the low amplitude phase of the 
modulation may indicate that these nonlinear effects are not so important in this phase of the modulation.}

\item{\cite{stothers} proposed that the enhanced convection lowers the pulsation amplitude in the small amplitude 
phase of the modulation, while the phase relation betwen the period changes and the amplitude variations depends 
on the physical parameters of the individual variable in a complex way. The pulsation period of MW Lyr is the 
longest about 1-2 days (0.1 phases of the modulation cycle) later than  the minimal amplitude phase, while it is 
the shortest about 3-4 days (0.2 phases of the modulation cycle) later than  the maximal amplitude phase.}

\item{The small variations in the phase differences of the lower order harmonic components of the light curves
during the Blazhko cycle indicate that the phase variations of these frequency components are very coherent.
If the modulation were caused by the interaction of close radial mode and nonradial mode frequencies \citep{dz}, 
then no such phase coherency would be expected.}

\item{The modulation of the light curve of MW~Lyr can be separated into amplitude and phase modulation components
using only one parameter, the phase of the $f_0$ pulsation frequency in each phase bins of the modulation. This is 
in agreement with the explanation of the Blazhko effect with period and amplitude changes as recently proposed 
by \citet{stothers}.}

\end{itemize}

It would also be important to know how common the detected properties of the light curve modulation of MW~Lyrae are. 
Only further similar observations of other Blazhko stars can give an answer to this question.

\section*{Acknowledgments}
We wish to thank the referee, Hiromoto Shibahashi for his useful comments that helped us to improve 
the paper. This research has made use of the SIMBAD database, operated at CDS Strasbourg, France. 
The financial support of OTKA grants K-68626 and T-048961 is acknowledged.
HAS thanks the US National Science Foundation for support under grants AST 0440061 and
AST 0607249.

\end{document}